\documentclass[preprint]{aastex}

\slugcomment{submitted to ApJ}

\def\uK{\mu {\rm K}}

\def\deg{^{\circ}}
\def\min{^{\prime}}
\def\fdg{\hbox{$.\!\!^\circ$}}

\def\fh{\hbox{$.\!\!^{\rm h}$}}

\def\lbra{{\langle}}
\def\rbra{{\rangle}}

\def\bn{{\bf n}}
\def\bu{{\bf u}}
\def\bv{{\bf v}}
\def\bw{{\bf w}}
\def\bx{{\bf x}}
\def\by{{\bf y}}
\def\bA{{\bf A}}
\def\bC{{\bf C}}
\def\bF{{\bf F}}
\def\bL{{\bf L}}
\def\bM{{\bf M}}
\def\bN{{\bf N}}
\def\bR{{\bf R}}
\def\bS{{\bf S}}
\def\bV{{\bf V}}
\def\Like{{\mathcal{L}}}
\def\pow{{\mathcal{C}}}
\def\tA{{\widetilde{A}}}
\def\tT{{\widetilde{T}}}

\def\tE{{\widetilde{E}}}
\def\tB{{\widetilde{B}}}
\def\tV{{\widetilde{V}}}
\def\tW{{\widetilde{W}}}
\def\tX{{\widetilde{X}}}
\def\tY{{\widetilde{Y}}}

\def\be{\begin{equation}}\def\bea{\begin{eqnarray}}\def\beaa{\begin{eqnarray*}}
\def\ee{\end{equation}}  \def\eea{\end{eqnarray}}  \def\eeaa{\end{eqnarray*}}

\tighten

\begin{document}

\title{Observational Strategies of CMB Temperature and Polarization \\
       Interferometry Experiments}

\author{Chan-Gyung Park\altaffilmark{1}, Kin-Wang Ng\altaffilmark{2,3}, 
        Changbom Park\altaffilmark{1}, Guo-Chin Liu\altaffilmark{3}, and
	Keiichi Umetsu\altaffilmark{3}} 
\altaffiltext{1}{Astronomy Program, School of Earth and Environmental Sciences,
                 Seoul National University, 151-742 Korea;
                 parkc@astro.snu.ac.kr, cbp@astro.snu.ac.kr.}
\altaffiltext{2}{Institute of Physics, Academia Sinica, Taipei, Taiwan 
                 11529, R.O.C.;
                 nkw@phys.sinica.edu.tw.}
\altaffiltext{3}{Institute of Astronomy and Astrophysics, Academia Sinica, 
                 Taipei, Taiwan 11529, R.O.C.; \\
		 gcliu@asiaa.sinica.edu.tw,
                 keiichi@asiaa.sinica.edu.tw.}

\begin{abstract}

We have simulated the interferometric observation of the Cosmic Microwave
Background (CMB) temperature and polarization fluctuations. 
We have constructed data pipelines from the time-ordered
raw visibility samples to the CMB power spectra which utilize the methods
of data compression, maximum likelihood analysis, and optimal subspace 
filtering. They are customized for three observational strategies, such as
the single pointing, the mosaicking, and the drift-scanning. 
For each strategy, derived are the optimal strategy parameters that yield 
band power estimates with minimum uncertainty. 
The results are general and can be applied to any close-packed array 
on a single platform such as the CBI and the forthcoming AMiBA experiments.

We have also studied the effect of rotation of the array platform
on the band power correlation by simulating the CBI single pointing 
observation. It is found that the band power anti-correlations can be reduced 
by rotating the platform and thus densely sampling the visibility plane.
This enables us to increase the resolution of the power spectrum 
in the $\ell$-space down to the limit of the sampling theorem 
($\Delta\ell = 226 \approx \pi / \theta$), which is narrower by a factor  
of about $\sqrt{2}$ than the resolution limit ($\Delta\ell \approx 300$) used 
in the recent CBI single pointing observation. The validity of this idea 
is demonstrated for a two-element interferometer that samples 
visibilities uniformly in the $uv$-annulus. 

From the fact that the visibilities are the Fourier modes of the CMB field
convolved with the beam, a fast unbiased estimator (FUE) of the CMB power 
spectra is developed and tested. 
It is shown that the FUE gives results very close to those from the quadratic
estimator method without requiring large computer resources even though 
uncertainties in the results increase.

\end{abstract}


\section{Introduction}

It is theoretically expected that the CMB is polarized. 
The CMB quadrupole anisotropy causes CMB photons polarized by Thomson 
scattering with electrons at the last scattering surface ($z \simeq 1,100$)
and during the reionization epoch ($z \lesssim 7$) (Hu \& White 1997).
The amplitude of polarization is predicted to be 1 -- 10\% of that of 
the temperature anisotropies, depending on angular scales.
The CMB polarization can provide useful information that is not much contained 
in the temperature anisotropy, such as the epoch of reionization or 
the tensor perturbations. 
Recently Keating et al. (2001) have reported the result of the POLAR 
experiment, giving upper limit of order 10 $\uK$ on the large-scale CMB 
polarization. The PIQUE experiment (Hedman et al. 2001, 2002) have also 
obtained a similar upper limit at subdegree scales. De Oliveira-Costa et al.
(2002) have tried to measure the cross-correlation $C^{TE}_{\ell}$ between 
temperature and $E$-mode polarization power spectra by cross-correlating
the PIQUE and Saskatoon data (Netterfield et al. 1997).
Many single dish and interferometry experiments such as MAXIPOL, POLAR, 
Polatron, COMPOSAR, CMB RoPE, DASI, and CBI are on-going to detect 
these faint polarized signals of CMB origin. 
Very recently, the DASI has reported a detection of the CMB $E$-mode 
polarization and the $TE$ cross-correlation by differencing the CMB 
fluctuations in two fields in 271 days of observation (Leitch et al. 2002b;
Kovac et al. 2002).

Since the Cambridge Anisotropy Telescope has detected the anisotropy 
on subdegree scales (CAT; O'Sullivan et al. 1995; Scott et al. 1996), 
the Degree Angular Scale Interferometer (DASI; Leitch et al. 2002a; Halverson 
et al. 2002), the Cosmic Background Imager 
(CBI\footnote{http://www.astro.caltech.edu/\~{}tjp/CBI}; Padin et al. 2001; 
Mason et al. 2002; Pearson et al. 2002), and the Very Small Array (VSA; 
Taylor et al. 2002; Scott et al. 2002) have also measured the CMB angular 
power spectrum down to subdegree scales.
A desirable feature of the interferometer for CMB observation is that 
it directly measures the power spectrum, and that the polarimetry is a 
routine. In addition, many systematic problems that are inherent 
in single-dish experiments, such as the ground and near field atmospheric 
pickup and spurious polarization signals, can be significantly reduced. 

The observation of CMB polarization using close-packed interferometers 
are also on-going or planned by the experiments like the DASI, CBI, 
and the forthcoming interim Array for Microwave Background Anisotropy 
(AMiBA\footnote{http://www.asiaa.sinica.edu.tw/amiba}; Lo et al. 2001).
They have full capabilities to probe the CMB temperature and polarization
simultaneously. The feed horns are able to detect $T$, $Q$, $U$, and $V$ 
Stokes parameters using complex linear or circular polarizers, 
aiming to detect CMB linear ($Q$ and $U$) polarization at wave numbers 
$100 \lesssim \ell \lesssim 4,000$.

This paper proposes the data analysis techniques for various strategies 
of the interferometric observation, especially when the $uv$-space beam size 
is larger than the structures of the CMB power spectrum in the $\ell$-space.
An attempt is made to increase the resolution of the estimated power
spectrum in the $\ell$-space. We have adopted three observational strategies,
namely single pointing, mosaicking, and drift-scanning methods, and tested
a few analysis methods extracting the angular power spectra from mock data
for efficiency.

The outline of this paper is as follows. In $\S2$ and $\S3$, beginning with 
a summary of CMB interferometric observation, we describe a theoretical 
formalism for analyzing CMB interferometric data. A prescription for making 
mock CMB observations with interferometric array is given. 
Application of each observational strategy is made in $\S4$. 
A fast unbiased power spectrum estimator is introduced in $\S5$. Finally,
the summary and discussion are given in $\S6$. 

\section{CMB Interferometry Experiment}

\subsection{Visibility}
Here we briefly summarize the CMB interferometry experiments 
(Hobson, Lasenby, \& Jones 1995; Hobson \& Maisinger 2002, hereafter HM02; 
White et al. 1999a, hereafter W99).
Only the observed CMB signal and instrument noise are considered, and  
other contaminations like the ground pickup or Galactic foregrounds are not
taken into account.   
The sky is assumed to be flat since the field size surveyed by 
the interferometric array is usually smaller than $10\deg$. 

The visibility function of CMB temperature fluctuations sampled at a pointing 
position $\by_m$ on the sky is the Fourier transform of the temperature 
anisotropy $\Delta T(\bx)$ on the sky weighted by the primary beam 
$A(\bx-\by_m,\nu)$, i.e.,
\be
    V_{\by_m}^{T}({\bu},\nu) = {\partial B_{\nu} \over \partial T}
         \int d^2 x A({\bx-\by_m, \nu}) \Delta T({\bx})
         e^{i 2\pi {\bu\cdot\bx}} 
\ee
(Hobson et al. 1995; W99). 
The factor $\partial B_{\nu} / \partial T \equiv (2k_B^3 T^2 / c^2 h^2)$ 
$x^4 e^x /(e^x -1)^2$ is the conversion factor from temperature to intensity, 
where $B_{\nu}(T)$ is the Planck function, $k_B$ is the Boltzmann's constant, 
$h$ is the Planck constant, $c$ is the speed of light, and 
$x \equiv h\nu/k_B T$. We assume $T \simeq T_{\rm CMB} = 2.725 \pm 0.002
~{\rm K}$ (Mather et al. 1999).
The {\it baseline vector} $\bu$ is a variable conjugate to the coordinate 
$\bx$, and has a dimension of the inverse angle.
In practice a vector connecting two centers of a pair of dishes determines 
a baseline vector $\bu$ in unit of observing wavelength. For a small field,
$u={1 \over 2} \cot (\pi/\ell) \approx \ell/2\pi$ (Hobson \& Magueijo 1996).
Each frequency channel gives an independent set of visibility samples with 
different baseline vectors.

The size of the primary beam $A(\bx)$ determines the area of the sky that 
is viewed and hence the size of the map while the maximum spacing of the array 
determines the resolution. $A(\bx)$ is normalized to unity at the pointing 
center, $A(0)=1$. 
The visibility in equation (1) can be also defined in the $uv$-domain,
\be
   V_{\by_m}^{T}({\bu},\nu)={\partial B_{\nu} \over \partial T} 
      \int d^2 w \tA(\bw,\nu) e^{i 2\pi {\bw\cdot\by_m}}
      \Delta \widetilde{T}(\bu-\bw),
\ee
where $\tA(\bu,\nu)$ and $\Delta\tT(\bu)$ are Fourier transform pairs
of $A(\bx,\nu)$ and $\Delta T(\bx)$, respectively. 
Let us define the generalized $uv$-space beam pattern $\tA_{\by_m}(\bu,\nu)$
at a pointing position $\by_m$ from the phase reference point,
\be
   \tA_{\by_m}(\bu,\nu) \equiv \tA(\bu,\nu) 
      e^{i 2\pi \bu\cdot\by_m}.
\ee
We adopt a beam pattern $\tA(\bu,\nu)$ given by W99, 
\be
   \tA(\bu,\nu) = {8 \over {\pi^2 D_\lambda}^2}
       \left( \arccos\left({u \over D_\lambda}\right) - {u \over D_\lambda} 
       \sqrt{1-\left({u \over D_\lambda}\right)^2} \right),
\ee
where $D_\lambda$ is the diameter of a dish ($D$) in unit of wavelength. 
It is defined by the autocorrelation of the electric field pattern of 
the flat-illuminated feed horn. This is equal to the autocorrelation of a 
pillbox of radius $D_{\lambda}/2$ (van Waerbeke 2000).
The primary beam pattern $A(\bx)$ obtained by the inverse Fourier transform 
of equation (4) is well approximated by a Gaussian function (for instance,
${\rm FWHM} \simeq 20\min$ for a dish with $D=60$ cm and the center frequency
$\nu_0 = 95$ GHz).

Visibilities of CMB $Q$ and $U$ Stokes parameters, $V_{\by_m}^{Q}$ and 
$V_{\by_m}^{U}$, can be similarly defined by replacing $\Delta T(\bx)$ 
with $Q(\bx)$ and $U(\bx)$, respectively.
The $Q$ and $U$ Stokes parameters depend on the observer's coordinate system. 
Under a rotation of coordinate system by $\psi$ on the sky, 
$Q \pm i U$ transforms as a spin-2 tensor, 
$(Q \pm iU) \rightarrow e^{\mp 2 i \psi} (Q \pm iU)$ (Zaldarriaga \& Seljak 
1997). It is useful to deal with rotationally invariant polarization 
quantities, known as $E$- and $B$-mode polarizations.
We follow the small-scale approximation of Seljak (1997) and Zaldarriaga \&
Seljak (1997) to describe CMB temperature and polarization fields,
\bea
   \Delta T(\bx) &=& \int d^2 u ~\Delta\tT(\bu) e^{-i 2\pi\bu \cdot \bx},
         \nonumber \\ 
   Q(\bx) &=& \int d^2 u ~[\tE(\bu) \cos 2\phi_{\bu} 
          - \tB(\bu) \sin 2\phi_{\bu} ] e^{-i 2\pi\bu \cdot \bx}, \nonumber \\
   U(\bx) &=& \int d^2 u ~[\tE(\bu) \sin 2\phi_{\bu} 
          + \tB(\bu) \cos 2\phi_{\bu} ] e^{-i 2\pi\bu \cdot \bx},
\eea
where $\tE(\bu)$ and $\tB(\bu)$ are the Fourier transforms of the $E$- and 
$B$-mode polarization fields, and $\phi_{\bu}$ is the direction angle of $\bu$.

\subsection{Visibility Covariance Matrix}

We are interested in measuring the angular power spectra of the CMB temperature
fluctuations ($\pow_{\ell}^{TT} \equiv \ell (\ell +1) C_{\ell}^{TT} / 2\pi$), 
polarizations ($\pow_{\ell}^{EE}$ and $\pow_{\ell}^{BB}$), 
and their cross-correlation ($\pow_{\ell}^{TE}$).
We can relate them with the power spectrum $S(u)$ defined in the $uv$-plane
under the flat-sky approximation by, for example, 
$\ell (\ell+1) C_{\ell}^{TE} / {2\pi}\approx 2\pi u^2 S_{TE}(u)$ 
where $S_{TE}(u)$ is defined by $\left< \Delta\tT(\bu) \tE^{*}(\bv) \right>
= S_{TE}(u) \delta(\bu-\bv)$, and likewise for other components. 

To estimate CMB temperature and polarization power spectra from the visibility
data using the maximum likelihood analysis, it is essential to construct 
visibility covariance matrices. Combining equations (2) and (5) gives 
the visibility covariance matrix, 
\bea
   M_{mn}^{ij} 
     &\equiv& \left< V_{\by_m}^X (\bu_i,\nu_i) V_{\by_n}^{Y*} 
         (\bu_j,\nu_j)\right>  \nonumber \\
     &=& {\partial B_{\nu_i} \over \partial T} {\partial B_{\nu_j} \over 
         \partial T} \int d^2 w \tA_{\by_m}(\bu_i - \bw,\nu_i) \tA_{\by_n}^{*}
         (\bu_j-\bw, \nu_j) \mathcal{S}_{XY}(\bw),
\eea
where $X$ and $Y$ correspond to $T$, $Q$, or $U$, and $i$ ($j$) is a visibility
data index at a pointing position $\by_m$ ($\by_n$).
All possible combinations of $\mathcal{S}_{XY}(\bw)$, defined by 
$\left< \tX(\bv) \tY^{*}(\bw) \right> = \mathcal{S}_{XY}(\bw) \delta(\bv-\bw)$, 
are listed in Table 1. 
We assume no correlation of temperature and $E$-mode polarization with 
$B$-mode, which is expected in common inflationary scenarios (Hu \& White 1997).
Changing the sign of $\bu_j$ in equation (6) gives another complex 
covariance matrix, 
\bea
   N_{mn}^{ij} 
     &\equiv& \left< V_{\by_m}^X (\bu_i,\nu_i) V_{\by_n}^{Y} 
              (\bu_j,\nu_j)\right> \nonumber \\
     &=& {\partial B_{\nu_i} \over \partial T} {\partial B_{\nu_j} \over 
      \partial T} \int d^2 w \tA_{\by_m}(\bu_i - \bw,\nu_i) \tA_{\by_n}
      (\bu_j+\bw, \nu_j) \mathcal{S}_{XY}(\bw).
\eea

In practice, it is more convenient to use real quantities rather than complex
ones. Separating the complex visibilities into real ($V^R$) and imaginary 
($V^I$) parts and combining $M_{mn}^{ij}$ and $N_{mn}^{ij}$ appropriately,
we obtain the followings (HM02, Appendix A),
\bea
   \lbra V_{\by_m}^{R} (\bu_i,\nu_i) V_{\by_n}^{R} (\bu_j,\nu_j) \rbra 
        &=& ( {\rm Re}[M_{mn}^{ij}] + {\rm Re}[N_{mn}^{ij}] )/2, 
        \nonumber \\
   \lbra V_{\by_m}^{I} (\bu_i,\nu_i) V_{\by_n}^{I} (\bu_j,\nu_j) \rbra 
        &=& ( {\rm Re}[M_{mn}^{ij}] - {\rm Re}[N_{mn}^{ij}] )/2, 
        \nonumber \\
   \lbra V_{\by_m}^{R} (\bu_i,\nu_i) V_{\by_n}^{I} (\bu_j,\nu_j) \rbra 
        &=& ( - {\rm Im}[M_{mn}^{ij}] + {\rm Im}[N_{mn}^{ij}] )/2, \nonumber \\
   \lbra V_{\by_m}^{I} (\bu_i,\nu_i) V_{\by_n}^{R} (\bu_j,\nu_j) \rbra 
        &=& ( {\rm Im}[M_{mn}^{ij}] + {\rm Im}[N_{mn}^{ij}] )/2. 
\eea

\subsection{Simulated Observations}
We use the CMBFAST power spectra $C_\ell^{TT}$, $C_\ell^{EE}$, and the 
cross-correlation $C_\ell^{TE}$ (Seljak \& Zaldarriaga 1996; Zaldarriaga,
Seljak, \& Bertschinger 1998) of a flat $\Lambda$CDM cosmological model.
The model parameters are $\Omega = 1$, $\Omega_\Lambda = 0.7$,
$h = 0.82$, $\Omega_b h^2 = 0.03$, and $n_s = 0.975$, which was given 
by the joint MAXIMA-BOOMERANG data analysis (Jaffe et al. 2001). 
We assume that the $B$-mode vanishes ($\tB(\bu) = 0$).
To generate $\Delta T$, $Q$ and $U$ fields on a patch of the sky we use 
the Fourier transform relations in equation (5). For a given $\ell$ 
we construct a $2 \times 2$ matrix $\bM$ that has $C_\ell^{TT}$ and 
$C_\ell^{EE}$ as the diagonal and $C_\ell^{TE}$ as the off-diagonal elements. 
We then Cholesky-decompose this matrix, $\bM = \bL \bL^T$, where $\bL$ is a 
lower triangular matrix, and assign random fluctuations $\Delta\tT(\bu)$ and 
$\tE(\bu)$ in each $uv$-cell using the following relation, 
\be
   \left( \begin{array}{c} \Delta\tT(\bu) \\ 
                          \tE(\bu)
          \end{array} \right) = \bL 
   \left( \begin{array}{c} r_1 \\
                           r_2
          \end{array} \right).
\ee
Here $r_1$ and $r_2$ are random complex numbers drawn from the Gaussian 
distribution with zero mean and unit variance (for definite normalization
of Eq. 9 in a finite field, see Wu 1999). 
The results are inverse Fourier transformed (Eq. 5) to yield 
$\Delta T(\bx)$, $Q(\bx)$, and $U(\bx)$.
These fields are then multiplied by the primary beam and Fourier transformed 
to give a regular array of visibilities (Eq. 1). An observation is simulated 
by sampling the regular array at the $\bu$ points specified by the 
dish configuration and observation strategy. The bilinear interpolation 
is used in the sampling.

The instrument noise is simulated by adding a random complex number to each 
visibility whose real and imaginary parts are drawn from a Gaussian 
distribution with the variance of the noise predicted in a real observation.
We use the sensitivity per baseline per polarization defined as 
(Wrobel \& Walker 1999; Ng 2001)
\be
   s_b = {1 \over \eta_s\eta_a} {{2k_B T_{sys}} \over {A_{phys}}} 
         {1 \over \sqrt{2 \Delta\nu}}, 
\ee
where $\Delta\nu$ is the bandwidth, $A_{phys}$ is the physical area of 
an elemental aperture, $T_{sys}$ is the system noise temperature, 
and $\eta_s$ ($\eta_a$) is the system (aperture) efficiency. We assume  
$\eta_s = \eta_a = 0.8$. Equation (10) is the sensitivity 
of a simple interferometer with a single real output consisting of the product 
of the voltages from two antennae. For a complex correlator, the noise
is statistically the same in each of the two channels and thus the real and
imaginary correlator outputs, $\sigma_r$ and $\sigma_i$ respectively, are 
equal with $\sigma_r = \sigma_i = \sigma_b$, where $\sigma_b = s_b / 
\sqrt{\tau_{acc}}$ and $\tau_{acc}$ is the correlator accumulation time.
If we use the simultaneous dual polarizer with noise level $\sigma_b$ per 
polarization, the $T$, $Q$, and $U$ visibilities will obey Gaussian statistics 
characterized by zero means and standard deviations, $\sigma_T = \sigma_Q = 
\sigma_U = \sigma_b / \sqrt{2}$.

The instrumental noise of a single dish experiment has a strong correlation 
with itself, for instance due to the $1/f$ noise statistics. 
But the noise of an interferometric array is randomly distributed over
the baselines and observations. This is because the noise in each correlator 
output of a pair of dishes has independent random noise statistics.
Therefore, the noise covariance matrix of the visibility data can be assumed 
to be diagonal, with the non-zero element simply given by the variance of 
the noise of each baseline component. 

\section{Data Analysis}

\subsection{Power Spectrum Estimation}

We summarize the basic scheme for the quadratic estimator of the CMB power 
spectrum based on the maximum likelihood analysis (Bond, Jaffe, \& Knox 1998; 
Tegmark 1997b; Tegmark \& de Oliveira-Costa 2001).  

We construct the data vector $\bV = (\bV^T, \bV^Q, \bV^U)$, with each 
visibility data point decomposed into real and imaginary parts. 
The observed visibility is the sum of contributions from the CMB signal 
and the instrument noise, $V_i = V_{Si}+ n_i$ ($i=1,2,\ldots,N_p$), 
where $N_p$ is the number of data points including all the real and imaginary 
components of $T$, $Q$, and $U$ visibilities. 
The covariance matrix of the visibility data becomes 
\be
   C_{ij} = \lbra V_i V_j \rbra = \lbra V_{Si} V_{Sj} \rbra
          + \lbra n_i n_j \rbra = S_{ij} + N_{ij}. 
\ee
The $S_{ij}$ and $N_{ij}$ are called signal and noise covariance matrices,
respectively, and the CMB signal is assumed to be uncorrelated with the 
instrument noise. The matrix element $S_{ij}$ is obtained from equation (8),
where we perform numerical integrations in equations (6) and (7). 
Assuming that the CMB $T$, $Q$, and $U$ fields are multivariate Gaussian
random variables, we get a maximum likelihood function,
\be
    \Like (\{ \pow_b \}) = {1 \over {(2\pi)^{N_p/2} |\bC|^{1/2}}}
          \exp\left(-\bV^T \bC^{-1} \bV /2\right)
\ee
for a given set of parameters 
$\{\pow_b\} = \{\pow_1^{TT},\ldots,\pow_{N_b/3+1}^{EE},\ldots,
\pow_{2 N_b/3+1}^{TE},\ldots,\pow_{N_b}^{TE}\}$.  Here the $\pow_b \equiv 
2\pi u_b^2 S(u_b)$ is an estimate of band power for a band centered on 
$u_b$ ($b = 1, 2, \ldots , N_b$), and $N_b$ is the total number of measured 
band powers. The superscript $T$ denotes transpose of a vector.
By using the Newton-Raphson method, after several iterations we can find 
a set of band powers that most likely fit the simulated data of the assumed
cosmological model and thus maximize the likelihood function.
The full quadratic estimator is
\bea
    \delta \pow_b &=& \sum_{b'} (\bF^{-1})_{bb'}  
         {{\partial \ln\Like} \over {\partial \pow_{b'}}} \nonumber \\
         &=& {1 \over 2} \sum_{b'} (\bF^{-1})_{bb'}
         {\rm Tr} \left[ (\bV\bV^T - \bC) \bC^{-1} {{\partial \bS} 
	                  \over {\partial\pow_{b'}}} \bC^{-1} \right], 
\eea
where the $N_b \times N_b$ matrix $\bF$ is the Fisher information matrix
defined as
\be
   F_{bb'} \equiv -\left< {{\partial^2 \ln \Like} \over {\partial \pow_b
         \partial \pow_{b'}}} \right>
         = {1 \over 2} {\rm Tr}\left( \bC^{-1}
           {{\partial \bS}
           \over
           {\partial \pow_b}} \bC^{-1} {{\partial \bS} \over
           {\partial \pow_{b'}}}
           \right).
\ee
The square root of the diagonal component of the inverse Fisher matrix,
$\left( \bF^{-1} \right)_{bb}^{1/2}$, is the minimum standard deviation 
of the measured band power $\pow_b$.
The derivative of the signal covariance matrix with respect to a band power
estimate, $\partial \bS / \partial \pow_b$, is obtained by combining 
$\partial M_{mn}^{ij} / \partial \pow_b$ and 
$\partial N_{mn}^{ij} / \partial \pow_b$ as in equation (8).
For a band power of temperature anisotropy ($\pow_b^{TT}$) with a band width 
ranging from $|\bu_{b_1}|$ to $|\bu_{b_2}|$, for example, 
\be
   {{\partial M_{mn}^{ij}} \over {\partial \pow_b^{TT}}}
             = {\partial B_{\nu_i} \over \partial T} 
               {\partial B_{\nu_j} \over \partial T}
                \int_{0}^{2\pi} {d\theta_w \over {2\pi}}
                \int_{|\bu_{b_1}|}^{|\bu_{b_2}|} {dw \over w}\tA_{\by_m}
                (\bu_i - \bw,\nu_i) \tA_{\by_n}^{*} (\bu_j-\bw, \nu_j),
\ee
and likewise for $\partial N_{mn}^{ij} / \partial \pow_b^{TT}$.
Cosine or sine factors are included in the integrand for band powers of 
polarization and cross-correlation ($\pow_b^{EE}$ and $\pow_b^{TE}$). 

For comparing a set of measured band powers $\{ \pow_b \}$ with cosmological 
models, we require a route to get band power expectation values 
$\left< \pow_b \right>$ from the model power spectrum $\pow_\ell$.
The expectation value for a band power spectrum estimate $\pow_b$ is given by 
\be
   \left< \pow_b \right> 
   = \sum_{b'} \left( \bF^{-1} \right)_{bb'} 
     \sum_{\ell\in{\rm all},~ \ell' \in b'} \pow_\ell F_{\ell\ell'}^s
   = \sum_{\ell} \left( W_\ell^b \Big/ \ell \right) \pow_\ell ,
\ee
where $\bF$ is the band-power Fisher matrix, and the $F_{\ell\ell'}^s$ is a
Fisher matrix element with bands of the individual multipoles.
$W_\ell^b / \ell$ is called the band power window function 
(Knox 1999; Pryke et al. 2002), and defined as
\be
   W_\ell^b \Big/ \ell
   \equiv \sum_{b'} \left( \bF^{-1} \right)_{bb'} 
          \sum_{\ell' \in b'} F_{\ell\ell'}^s , ~{\rm and}~ 
   \sum_{\ell} W_\ell^b / \ell = 1.
\ee
This function can be considered as a filter that averages out the power
spectrum with variance weighting scheme (see Tegmark 1997b; Tegmark \& 
de Oliveira-Costa 2001 for a related discussion).

Although the quadratic estimator is much faster than the direct evaluation
of the likelihood function in finding the maxima, it still demands intensive
computing power. The most time consuming part is the calculation of $\bF$, 
which is $\mathcal{O}(N_p^3)$ operation.
The visibility data obtained from mosaicking or drift-scanning usually
involves a large amount of visibilities, but many of them are strongly 
correlated.
One way to increase the efficiency of the calculation is to work with 
a subset of the data that contains most of the signal. This transformation 
can be obtained using the {\it optimal subspace filtering}, also known as 
the signal-to-noise eigenmode transform (see Tegmark, Taylor, \& Heavens 1997 
for a review). 
In this paper we follow the transformation method given by Bond et al. 
(1998, Appendix A). We first perform a whitening transformation,
\bea
   \bN &\longrightarrow& \bN^{-1/2} \bN \bN^{-1/2} = {\bf I},  \nonumber \\
   \bS &\longrightarrow& \bN^{-1/2} \bS \bN^{-1/2},   \\
   \bV &\longrightarrow& \bN^{-1/2} \bV,  \nonumber
\eea
where the whitening filter $\bN^{-1/2}$ is the inverse of the Cholesky
decomposition of $\bN$. 
Since $\bN$ is diagonal, the above quantities reduce to simple forms,
$\delta_{ij}$, $S_{ij}/\sigma_i \sigma_j$, and $V_i /\sigma_i$, respectively. 
By diagonalizing the signal covariance matrix using the similarity 
transformation, we get 
\bea
     \bS &\longrightarrow& \bR^T (\bN^{-1/2}\bS\bN^{-1/2}) \bR 
           ={\mbox{\boldmath $\varepsilon$}}
           ={\rm diag}(\varepsilon_k),  \nonumber \\
\bV &\longrightarrow& \bR^T (\bN^{-1/2} \bV) = \bV^{\varepsilon},
\eea
where diagonal components of ${\mbox {\boldmath $\varepsilon$}}$ are 
eigenvalues of the whitened signal covariance matrix, and $\bR$ is an unitary
matrix of which the $k$th column vector is an eigenvector corresponding 
to the $k$th eigenvalue $\varepsilon_k$. 
Transformation of $\partial \bS / \partial \pow_b$ into the eigenbasis can be 
also performed in a similar way. In this new basis, the data 
$V_k^{\varepsilon}$'s are uncorrelated variables with variances 
$\left< V_k^{\varepsilon 2} \right> = 1 + \varepsilon_k$.
The $\varepsilon_k$'s are usually called the signal-to-noise eigenmodes.
In our analysis, we can keep only the top 20 -- 30\% of the modes and 
treat them as a new data set. 
This is an excellent approximation to the original dataset because 
the contribution of eigenmodes with lower signal-to-noise ratios to the CMB
signal is relatively negligible.
By replacing the covariance matrix and the data vector in equations (13) and 
(14) with the transformed ones in equation (19), we get a new version of 
quadratic estimator in this eigenmode basis 
(see Eqs. A10 and A11 of Bond et al. 1998).
Although in this formalism the transformation into the signal-to-noise basis,
which is $\mathcal{O}(N_p^3)$ operation, is the most expensive part, 
the remaining procedures do not require large computation. 
  
\subsection{Data Compression: Pixelization in real- and ${\mbox{\boldmath 
$uv$}}$-spaces}

It is essential to compress the visibility data, since the quadratic 
estimator for power spectrum requires very large operation proportional to 
$\mathcal{O}(N_p^3)$. The optimal subspace filtering described in $\S3.1$
is one method for data compression. For a completely close-packed array, 
we can reduce the length of the data set by averaging all the visibilities 
with the same baseline vector for each component ($T$, $Q$, $U$) and 
observing frequency. 
For example, the 7-element AMiBA in hexagonal configuration gives 21 baselines 
for each frequency channel, and they can be reduced to 9 independent baselines
(Fig. 1$a$). 

In many CMB experiments, a telescope scans the sky continuously many times 
over a specified survey area, and generates the time-ordered data stream. 
Analyzing this raw data directly to estimate cosmological parameters is 
intractable. The same problem arises in the drift-scanning mode of the CMB
interferometric observation (see $\S4.3$). Furthermore, if the platform 
of the interferometer rotates with respect to the survey field during 
the observation we will get a even larger visibility data set from 
a denser coverage of the $uv$-plane with the moving baseline vectors. 
In this case pixelization in the real- and/or $uv$-spaces can serve 
as a data compression method (Wu 2002). 
HM02 have developed a pixelization scheme in the $uv$-plane (see also Myers 
et al. 2002). This can be understood as a commonly-used map-making process 
(Tegmark 1997a; Stompor et al. 2002).   
Here we generalize the $uv$-space pixelization method of HM02 to include 
the real-space pixelization as well. 

The time-ordered visibility data obtained by an interferometric array can be 
modeled by the pixelized visibility data vector $\bV^{\rm pix}$ multiplied 
by a pointing matrix $\bA$ plus a time-ordered instrumental noise vector $\bn$, 
\be
    \bV^{\rm tod} = \bA \bV^{\rm pix} + \bn . 
\ee
In component notation it is
\be
   V_{\by(t_i)}^{\rm tod}(\bu_k) = \sum_{p,l}A_{(ik)(pl)}V_{\by_p}^{\rm pix}
      (\bu_l) + n_{\by(t_i)}(\bu_k),   
\ee
where $\by(t_i)$ and $\bu_k$ are the observing position and the baseline vector
at time $t_i$, and $\by_p$ and $\bu_l$ are the pixelized real- and $uv$-space
positions, respectively.
The pointing matrix $\bA$ is determined entirely by the observational or 
scanning scheme. Suppose that the total number of time-ordered visibility data
at a frequency $\nu$ is $N_{\rm tod}$, and the length of the pixelized 
visibility is $N_{\rm pix}$. Then the $N_{\rm tod} \times N_{\rm pix}$ 
pointing matrix $\bA$ is defined with matrix elements $A_{(ik)(pl)}=1$ 
if $(\by(t_i), \bu_k)$ lies in the cell specified by $(\by_p, \bu_l)$, 
otherwise $A_{(ik)(pl)}=0$. 

The noise covariance matrix of the time-ordered visibility data can be taken
to be diagonal (see $\S2.3$),
\be
   (\bN_{\rm t})_{(ik)(i' k')} 
      = \left< n_{\by(t_i)}(\bu_k) n_{\by(t_{i'})}(\bu_{k'}) \right>
      = \sigma_{ik}^2 \delta_{ii'} \delta_{kk'}.
\ee
A solution ${\bf \tV}^{\rm pix}$ of equation (20) can be obtained by applying 
appropriate filtering matrix ${\bf W}$, i.e.,
\be
   {\bf \tV}^{\rm pix} = {\bf W} \bV^{\rm tod}.
\ee
We adopt the well-known minimum-variance filter (Janssen \& Gulkis 1992; 
Tegmark 1997a), 
\be
   {\bf W} = \left( \bA^T \bN_{\rm t}^{-1} \bA \right)^{-1} \bA^T 
             \bN_{\rm t}^{-1}.
\ee
This filter has an attractive property that the reconstruction error, 
$\delta\bV_e \equiv {\bf\tV}^{\rm pix} - {\bV}^{\rm pix} 
= (\bA^T \bN_{\rm t}^{-1} \bA)^{-1}$ $\bA^T \bN_{\rm t}^{-1}\bn$,
becomes independent of $\bV^{\rm pix}$. The noise covariance matrix in the
pixelized visibility data becomes 
\be
   \bN_{\rm p} \equiv \left< \delta\bV_e \delta\bV_e^T \right> 
      = \left( \bA^T \bN_{\rm t}^{-1} \bA \right)^{-1}.
\ee

Generally, the reconstructed visibility ${\bf\tV^{\rm pix}}$ is the 
minimum-variance estimate of $\bV^{\rm pix}$, and becomes maximum-likelihood
estimate provided that the probability distribution of $\bn$ is Gaussian. 
Since the time-ordered noise covariance matrix $\bN_{\rm t}$ is diagonal, 
the matrix equations for the pixelized visibility and its noise covariance 
(Eqs. 23 and 25) are simple bin-averaging processes, i.e.,
\be
   \tV_{\by_p}^{\rm pix}(\bu_l) = {
      {\sum_{i \in p, k \in l} V_{\by(t_i)}^{\rm tod}(\bu_k)/\sigma_{ik}^2 } 
      \over 
      {\sum_{i \in p, k \in l} 1/\sigma_{ik}^2}
   },
\ee
and
\be
   (\bN_{\rm p})_{(pl)(p'l')} 
       =  \Big\{ 1\Big/\sum_{i \in p, k \in l} 
          1/\sigma_{ik}^2 \Big\} \delta_{pp'} \delta_{ll'}.
\ee
From equations (23) and (24) it can be understood that when ${\bf W}$ acts 
on $\bV^{\rm tod}$, the first operation $\bA^T \bN_{\rm t}^{-1} \bV^{\rm tod}$ 
maps the raw data into the pixelized visibility space; this step averages 
all data with coordinates $\left( \by(t_i), \bu_k \right)$ that belong 
to the cell $\left( \by_p, \bu_l \right)$, weighed with noise variance 
$\sigma_{ik}^2$. 
The second one, $\left( \bA^T \bN_{\rm t}^{-1} \bA \right)^{-1}$, is simply 
the normalization factor.
 
The remaining issue is to determine the size of the pixel (both in real- and 
$uv$-spaces) before processing equations (26) and (27).
In virtue of the sampling theorem the size of the $uv$-pixel, 
$\Delta u_{\rm pix}$, must not be bigger than a half of the FWHM of the beam, 
i.e., $\Delta u_{\rm pix} \lesssim \Delta u_{\rm fwhm}/2$ (Hobson et al. 1995; 
HM02).  For single pointing observation, the half-width of the beam defined 
in equation (4) is given by 
\be
   \Delta u_{\rm fwhm}/2 \simeq 12 
        \left({D \over {30~{\rm cm}}}\right)
	\left({\nu \over {30~{\rm GHz}}}\right).
\ee
For example, $\Delta u_{\rm fwhm} /2  \simeq 36$ for the CBI ($D=90$ cm, 
$\nu_0\simeq 30$ GHz) and $\Delta u_{\rm fwhm} /2 \simeq 77$ for the AMiBA 
($D=60$ cm, $\nu_0 = 95$ GHz; Fig. 1$b$). 
We can set a similar limit for the real-space pixel size, i.e., 
$\Delta \theta_{\rm p} \lesssim \Delta \theta_{\rm fwhm}/2$. 

\section{Observational Strategies and Applications: CBI and AMiBA Examples}

In this section we describe three observational strategies and give 
simulation examples for the CBI and the AMiBA. The forthcoming interim 
AMiBA is an interferometric array of 7 elements. 
The characteristics of the AMiBA is summarized in Figure 1. 
Previous theoretical works and simulations on the AMiBA project can be found 
in Ng (2001, 2002), Pen et al. (2002), and Park \& Park (2002). 

\subsection{Single Pointing Observation}

In CMB interferometric observation, the observational strategy depends on 
the characteristics of the interferometric array, especially on the dish 
configuration and on the primary beam size in the $uv$-plane. 
In the {\it single pointing} strategy, the array observes many points 
on the sky. This strategy is appropriate for the interferometer that has 
uniform $uv$-coverage and sufficiently small primary $uv$-beam size.
This interferometer like DASI can cover a wide range of angular scales 
and resolve the structures in the CMB power spectrum. 
One can increase the signal-to-noise ratio of visibility 
and reduce the sample variance of the CMB fluctuations by increasing 
integration time per field ($t_{\rm f}$) and the number of independent fields 
($n_{\rm f}$), respectively (see Halverson et al. 2002 for the recent results
of the single pointing observation of DASI).

In Figure 2$a$ we plot the CMB power spectra of temperature anisotropy 
measured from the CBI mock survey adopting the single pointing strategy 
($n_{\rm f}=30$, $t_{\rm f}=72$ hours) with band power widths of 
$\Delta u = \Delta u_{\rm fwhm}/2 = 36$ (filled circles) and 72 (open circles).
We assume that 13 elemental apertures ($D=90$ cm) are located on a platform 
with DASI configuration given by White et al. (1999b, Table 1). 
We also assume that CBI complex correlators ($T_{sys}=30$ K) have 5 channels
around 26 -- 36 GHz.
During the observation of a specified field, the platform does not
change its parallactic angle so that all baseline vectors keep constant 
orientations relative to the field.
To estimate band powers from the mock visibility samples, we used the quadratic
estimator described in $\S3.1$.
At high $\ell \gtrsim 2,000$, where noise dominates, strong band power 
anti-correlations exist. We define the band power correlation by
$M_{bb'}/\sqrt{M_{bb}M_{b'b'}}$, where $M_{bb'}$ is the inverse Fisher 
matrix element $(\bF^{-1})_{bb'}$.
Correlations between neighboring bands at lower $\ell \lesssim 2,000$ 
are from $-57$\% to $-35$\% for $\Delta u = 36$, 
while they are between $-19$\% and $-16$\% for $\Delta u = 72$.

We can reduce the correlations among the band power estimates by sampling 
the visibilities in the $uv$-plane in such a way that the $uv$-space is 
better covered both in radial and transverse directions
and the window functions of various baselines are densely overlapping.
As a result of this, the resolution of the power spectrum in the $\ell$-space
(hereafter $\Delta\ell$-resolution) can be increased.
To demonstrate this idea, we consider a case with the CBI array whose platform 
rotates by $60\deg$ with steps of $10\deg$, $6\deg$, and $2\deg$ during 
the observation of a single field to get visibility samples with 6, 10, and 
30 different orientations, respectively. These situations are very similar 
to the CBI deep fields case (Mason et al. 2002). 
Furthermore, the visibility data set with 6 orientations approximately 
equals to the data set in each pointing in the recent CBI mosaic observation 
(Pearson et al. 2002) that has 78 baselines, 10 channels, and 42 mosaic 
pointings. Here the total number of visibilities is $N_v \approx 200,000$,
and the effective number of orientations is $200,000/780/42 \simeq 6$.
We pixelize the $uv$-plane with $\Delta u_{\rm pix} = 12$, and use equations 
(26) and (27) to reduce the data size.
Figure 2$b$ shows the power spectra measured from the pixelized CBI mock
visibilities with 6 ($n_{\rm f}=15$, $t_{\rm f}=24^{\rm h}\times 6$;
open triangles), 10 ($n_{\rm f}=9$, $t_{\rm f}=24^{\rm h}\times 10$;
filled circles), and 30 ($n_{\rm f}=6$, $t_{\rm f}=12^{\rm h}\times 30$; 
open circles) orientations. 
The anti-correlations between the power estimates at $\ell < 2,000$ are 
significantly reduced as the number of orientations increases. 
Table 2 summarizes the band power correlations and the band power uncertainties
in the power estimations of the above four cases.

To further investigate the above method of increasing $\Delta\ell$-resolution
via the {\it uv-mosaicking}, we have simulated mock observations using 
a two-element interferometer with $D=20$ cm ($\Delta u_{\rm fwhm}=51.2$) 
and two frequency channels (20 GHz whole bandwidth centered at $\nu_0 = 95$ 
GHz). 
The dishes change their separation from 40 cm to 60 cm, during which 
the platform is rotated to obtain nearly uniform visibility samples. 
We adopt different noise levels depending on the amplitudes of the model 
power spectrum, i.e., $\sigma_b = 43 (C_\ell / C_{600})^{1/2}$ mJy  
per visibility.
Figure 3 shows the distribution of visibilities in the $uv$-plane 
(upper right box) and the band powers estimated from 30 mock observations.
The band width of the intermediate five bands is $\Delta u=21$ 
($\Delta\ell = 132$), and the band power correlations are between 
$-29$\% and $-18$\%, with an average of $-24$\%.

Hobson \& Magueijo (1996) and Tegmark (1997c) derived useful formulae
for the uncertainties of the measured power spectrum. 
Suppose that $\bar{S}(u_b)$ denotes the power $S(u)$ averaged over a band
ranging from $u_{b_1}$ to $u_{b_2}$ centered on $u_b$. 
The rms error for this band power is given by 
$\sqrt{2} \left(\bar{S}(u_b) + \bar{S}^{\rm noise}(u_b)\right)$, 
where the first part is due to the sample variance, and the second is 
due to the noise variance.
If there are $N(u_b)$ independent eigenmodes in a band and if $n_{\rm f}$
independent fields are observed, the error will drop by a factor of 
$\sqrt{n_{\rm f} N(u_b)}$. 

Let us grid the $uv$-plane with cell size $u_0$. In the single pointing 
strategy, the natural choice for the cell size is 
$u_0 \simeq 1/\sqrt{\Omega_s}$ where $\Omega_s$ is the solid angle of 
the primary beam. Then, the noise power per $uv$-cell is given by
\be
   \sigma_N^2 = {{s^2 \Omega_s^2} \over {n_{\rm vis} t_{\rm f} }}
              = {{s^2 \Omega_s^3} \over {\rho_{\rm vis} t_{\rm f}}},
\ee
where $s$ is the sensitivity of the instrument in units of 
$\uK$ ${\rm sec}^{1/2}$ given by 
$s = \left( \partial B_\nu / \partial T \right)^{-1} s_b / \Omega_s$,
$n_{\rm vis}$ is the number of visibilities per $uv$-cell, and 
$\rho_{\rm vis}$ is the number of visibilities per unit area in the $uv$-plane;
$n_{\rm vis} = \rho_{\rm vis} u_0^2$.
Suppose that the sampled area in the half of the $uv$-plane is 
$A_0 = \pi(u_{b_2}^2 - u_{b_1}^2)/2$, and $N_v$ is the number of visibilities 
in this area ($N_v = A_0 \rho_{\rm vis}$). 
Given the detector sensitivity $s$, $uv$-plane coverage $A_0$, and total 
observation time $t_{\rm tot}$, we can specify a quantity to keep constant,
\be
   w^{-1} = {{s^2 A_0} \over {t_{\rm tot}}}
          = {{\sigma_N^2 N_v} \over {\Omega_s^3 n_{\rm f}}},
\ee
where $w$ can be considered as the noise weight per unit $uv$-area,
and $t_{\rm tot} = n_{\rm f} t_{\rm f}$.

From equations (29) and (30) the noise power spectrum can be obtained, 
$\bar{S}^{\rm noise}(u_b) \simeq \sigma_N^2 u_0^2 
= w^{-1} n_{\rm f} \Omega_s^2 / N_v$.
If $A_{\rm eff}(u_b)$ is the effective area occupied by $N_v^i$
independent visibility components, then 
$N(u_b) \approx N_v^i = A_{\rm eff}(u_b) \Omega_s$ for sufficiently 
wide band. Therefore, the uncertainty limit for the band power estimation 
becomes
\be
   \Delta \bar{S}(u_b) \approx 
	    \sqrt{{2} \over {n_{\rm f}A_{\rm eff}(u_b) \Omega_s}}
	    \left( \bar{S}(u_b) + {{w^{-1} n_{\rm f}\Omega_s^2 }
            \over {N_v}} \right).
\ee
How many pointings do we need to make to minimize the band power uncertainty? 
With $t_{\rm tot}$ and $\Omega_s$ fixed, we minimize $\Delta\bar{S}(u_b)$ 
with respect to $n_{\rm f}$ to get the best choice
\be
   n_{\rm f} = {{w N_v \bar{S}(u_b)} \over {\Omega_s^2}}
             = {{t_{\rm tot} N_v \bar{S}(u_b)} \over {s^2 A_0 \Omega_s^2}}.
\ee
If $n_{\rm f}$ is fixed instead, the optimal observation time for each pointing
becomes $t_{\rm f} = s^2 A_0 \Omega_s^2 / N_v \bar{S}(u_b)$.
This corresponds to the limit making the noise and the sample variance
contributions equal.

Let us consider an example using a two-element AMiBA with a single frequency 
channel ($N_v = N_v^i = 1$). 
We measure only one $E$-mode band power with a wide band centered 
at $u_b \simeq 206$, with $u_{b_1} \simeq 129$ and $u_{b_2} \simeq 283$, 
given by the FWHM limit of the $uv$-beam pattern (Eq. 28).  
The fluctuation power of $E$-polarization is expected to be 
$\pow_b^{EE} \approx 20$ $\uK^2$ in our $\Lambda$CDM model.
With $t_{\rm tot} = 6$ months, we obtain the optimal parameters, 
$n_{\rm f} = 32$ and $t_{\rm f}=135$ hours, which give the band power 
uncertainty $\Delta \pow_b \equiv 2\pi u_b^2 \Delta \bar{S}(u_b) \simeq 10$ 
$\uK^2$. Our mock observations with these parameters has confirmed 
the validity of equation (31).

The time required for a $4\sigma$ detection of the CMB polarization using 
the AMiBA with 7 dishes is worth discussing. 
Unless we increase the number of independent modes $N(u_b)$ within the band, 
it would be impossible to get the $4\sigma$ limit by observing only one field 
even with infinite integration time due to the sample variance. 
Here we consider only the shortest baseline components ($N_v=12$, $N_v^i=3$). 
From equation (31) with a condition $\Delta\bar{S}(u_b)=\bar{S}(u_b)/4$,
a $4\sigma$ detection of the $E$-polarization can be achieved in a minimum 
total observation time of $t_{\rm tot}=20$ days when $n_{\rm f}=43$.

\subsection{Mosaicking}

The resolution in the $uv$-space is limited by the area of the sky that
is surveyed, which is equal to the size of the primary beam in a single
pointing observation. By {\it mosaicking} several contiguous pointing 
observations, we can increase $\Delta\ell$-resolution in the band power
estimates. Mosaicking does not increase the $u$-range, but simply enhances 
the resolution by allowing us to follow more periods of a given wave (see W99 
for a complete discussion). 
For close-packed interferometers such as the AMiBA and the CBI whose $uv$-beam
size larger than the structure of the CMB power spectrum, 
mosaicking is essential to increase the $\Delta \ell$-resolution. 
In particular, since the 7-element AMiBA has baselines of only three different 
lengths (see Fig. 1$c$), the $uv$-coverage is so sparse that the band powers 
estimated from a single pointing observation will have large sample variances
and strong anti-correlations (Ng 2001).

The increase of $\Delta\ell$-resolution by mosaicking can be demonstrated
by using the effective $uv$-beam pattern 
\be
   \tA_{\rm eff}(\bu,\nu) 
   \equiv {1 \over N_{\rm mo}} \sum_{i=1}^{N_{\rm mo}} \tA_{\by_i}(\bu,\nu)
   = {1 \over N_{\rm mo}} \sum_{i=1}^{N_{\rm mo}} 
     \tA(\bu,\nu) e^{i 2\pi\bu\cdot\by_i},
\ee
where $N_{\rm mo}$ is the total number of contiguous pointings and $\by_i$ 
is the $i$th pointing position on the sky. 
Figure 4 shows the profiles of the effective $uv$-beams resulted from 
7- and 19-pointing mosaics with separation $\delta\theta_{\rm mo}$ in a 
hexagonal configuration. As the separation $\delta\theta_{\rm mo}$ increases, 
the effective width becomes narrower. 
However, for $\delta\theta_{\rm mo} > 20 \min$, complex correlations among 
the band power estimates are expected due to the sidelobe effects 
(e.g., see the long dashed curve in Fig. 4$b$).
Figure 5 shows an example of the CMB temperature and polarization power 
spectra estimates expected in the AMiBA 19-pointing hexagonal mosaic mock 
observation in a period of 6 months.
Using the AMiBA configured as in the Figure 1, we have made 10 independent
mosaics with $\delta\theta_{\rm mo}=15\arcmin$ and an integration time of 22.7 
hours per pointing ($n_{\rm f}=10$, $t_{\rm f}=22\fh7 \times 19$). 
The quadratic estimator and the optimal subspace filtering are applied 
to estimate the band powers with $\Delta u = 90$ (see $\S3.1$).
The band power window functions and the corresponding band power expectation 
values $\left< \pow_b \right>$ calculated from equations (16) and (17) 
are also shown. It shows that the measured band powers excellently match 
with the theoretical values.

For the mosaicking strategy, the uncertainty limit of the band power 
estimation in equation (31) is modified to
\be
 \Delta \bar{S}(u_b) 
        \approx \sqrt{{2} \over {n_{\rm f}A_{\rm eff}(u_b) \Omega_{\rm mo}}}
        \left( \bar{S}(u_b) + {{w^{-1} n_{\rm f}\Omega_s \Omega_{\rm mo} } 
        \over {N_v}} \right),
\ee
where $\Omega_{\rm mo}$ is the mosaicked area of each field. Here we have 
adopted a new $uv$-pixel size $u_0^2 \simeq 1 / \Omega_{\rm mo}$,
and a new noise power per $uv$-cell 
$\sigma_N^2 =  s^2 \Omega_s \Omega_{\rm mo} / n_{\rm vis} t_{\rm f}$.
The optimal mosaicked area can be obtained by minimizing $\Delta\bar{S}(u_b)$
with respect to $\Omega_{\rm mo}$, giving the condition,
$\bar{S}(u_b) = w^{-1} n_{\rm f} \Omega_s \Omega_{\rm mo} / N_v$.
Given a total observation time $t_{\rm tot}=n_{\rm f} t_{\rm f}$, 
the best choice for the mosaicked area becomes
\be
   \Omega_{\rm mo} = {{w N_v} \over {n_{\rm f} \Omega_s}} \bar{S}(u_b)
            = {{t_{\rm f} N_v} \over {s^2 A_0 \Omega_s}} \bar{S}(u_b). 
\ee

As an example, let us consider the visibilities measured by the shortest 
baselines in the AMiBA polarization observation, as described in $\S4.1$.
With 7 dishes, the total number of visibility outputs is $N_v = 12$
with 3 independent baselines ($N_v^i = 3$), and the number of independent 
modes within the wide band ($u_b \simeq 206$, $u \simeq$ 129 -- 283) 
becomes $N(u_b) \approx N_v^i (\Omega_{\rm mo} / \Omega_s) 
= A_{\rm eff}(u_b) \Omega_{\rm mo}$. 
From equation (35), we obtain an optimal square-shaped mosaic with size of
$\theta_{\rm opt}$, assuming a total observation time $t_{\rm tot} = 6$ months,
\be
     \theta_{\rm opt} = \Omega_{\rm mo}^{1/2} \approx { {7\deg} 
                        \over {\sqrt{n_{\rm f}}}   }.
\ee
Hence, from equations (34)--(36), the uncertainty of the band power 
estimate only depends on the total survey area $n_{\rm f}\Omega_{\rm mo}$ 
for a given total observation time $t_{\rm tot}$, and is expected to be
$\Delta\pow_b \approx 2$ $\uK^2$ for $t_{\rm tot} = 6$ months. 
Note that choosing a smaller band width will increase $\Delta\pow_b$; 
if there are $n_b$ subbands within a certain wide band, the size of error bar 
for each subband in the wide band would be roughly $\Delta\pow_b n_b^{1/2}$.
Increasing $\Omega_{\rm mo}$ (or decreasing $n_{\rm f}$) will give 
better results since we can get more information from larger scales and 
hence increase the resolution of the band power width 
$\Delta \ell \gtrsim \pi / \theta_{\rm opt}$. 
Table 3 summarizes the optimal parameter choices for the single pointing
and mosaicking strategies, given a total observation time $t_{\rm tot}=6$ 
months.

Figures 6$a$, 6$b$, and 6$c$ show the CMB power spectra expected in an AMiBA 
experiment when it performs $12 \times 12$ mosaic 
($\delta\theta_{\rm mo}=15\arcmin$) observations of $3\deg \times 3\deg$ area
over five fields ($n_{\rm f}=5$) with 6 hours of integration time per pointing.
We have assumed two frequency channels and used the quadratic estimator as well
as the signal-to-noise eigenmode analysis in our computation. 
Except for the first and the last bands, the band power width is chosen to be 
$\Delta u=45$ ($\Delta\ell = 283$).
The $\ell$-location of each band power is roughly estimated from 
the noise-weighted window function defined as 
\be
    W_\ell^{N} \Big/ \ell \equiv {\rm Tr}\left[\bN^{-1/2} 
               {{\partial \bS} \over {\partial \pow_\ell}} 
	       \bN^{-1/2}\right] \Big/N_\ell,
\ee
where the prewhitening transformation in $\S3.1$ is applied,
and $N_\ell$ is the number of data points that contribute to the sensitivity
at $\ell$ (Bond et al. 1998). Note that equation (37) is an approximate 
window function, and the Fisher matrix-derived window function (Eq. 17) 
should be used in general.
The band power correlations are sufficiently small with the average of about
$-10$\%. We have made two sets of measurement of the band powers with band 
centers shifted by the half of the band width with respect to each other. 
The $\Lambda$CDM model power spectra convolved with a box of band power width
weighted by the window function in equation (37) are plotted to guide 
comparison between the measured powers and the theory (thin curves and squares).
Since there are about 4 bands ($n_b \approx 4$) within the range of a
wide band in this case ($u\simeq$ 129 -- 283 centered at $u_b \simeq 206$), 
one would expect the estimated $E$-polarization band powers near $u_b=206$ 
to be $\sigma(\pow_b) \approx \Delta\pow_b n_b^{1/2} \approx 4$ $\uK^2$, 
which coincides with the uncertainty of the fourth band power in Figure 6$b$ 
($4.1$ $\uK^2$) and validates equation (34). 

\subsection{Drift-Scanning}

In the mosaic strategy, one has to track every pointing of a specified mosaic 
field for a sufficiently long time. This inevitably suffers from severe 
ground pickups since the ground emission usually changes abruptly
as the platform of the array turns around during its tracking.
{\it Drift-scanning} is the simplest way to remove the ground contamination
because an interferometer is insensitive to the ground emission that is 
a DC signal (Pen et al. 2002). In this strategy the direction of the array 
is fixed while the sky drifts along the constant declination as the earth 
rotates. 
After some time, the array slews to the original starting position, and 
observes the sky along the same scanning path.
In the drift-scan mode, the visibility is a function of time along a scan path.

To estimate the power spectrum by using reasonable computer resources,
the time-ordered visibility samples must be compressed (see $\S3.2$).
During the compression, the data experiences effects of smoothing over 
the pixel scale.  
If we consider only the real-space pixelization, a pixelized visibility 
$V_{\by_p}^{T}(\bu_k)$ is related with the time-ordered visibilities 
$V_{\by(t)}^{T}(\bu_k)$ by
\bea
   V_{\by_p}^{T}(\bu_k) &=& \int d^2 y W_{\rm p}(\by - \by_p) 
                             V_{\by(t)}^{T}(\bu_k) \nonumber \\
      &=& {\partial B_{\nu} \over \partial T}
          \int d^2 y W_{\rm p}(\by-\by_p) \int d^2 v
          \tA(\bv) e^{i 2\pi \bv \cdot \by} 
	  \Delta\tT(\bu_k - \bv) \nonumber \\
      &=& {\partial B_{\nu} \over \partial T}
          \int d^2 v \tA(\bv) \tW_{\rm p}(\bv)
          e^{i 2\pi \bv \cdot \by_p} 
	  \Delta\tT(\bu_k - \bv),
\eea
where $\tW_{\rm p}(\bv)$ is the Fourier transform of the smoothing
filter $W_{\rm p}(\by)$, $\by=\by(t)$ is the drift-scan path, and $\by_p$ 
denotes the real-space pixel position on the sky.

We have made an idealized mock AMiBA experiment in which the array drift-scans
a small square field of $2\deg \times 2\deg$ size near the celestial equator
(for flat sky approximation). The platform does not change its parallactic 
angle during the observation. This enables us to consider only the real-space
pixelization in the data compression. 
After scanning a $2\deg$ path, the next path $10\min$ apart from the previous
one is scanned. The $T$, $Q$, and $U$ visibilities are functions of time 
determined by the rotating speed of the sky, and are accumulated every 
8 seconds. Also, according to equation (36), the AMiBA drift-scans a square 
field of $2\deg \times 2\deg$ many times in a way described above for 
$t_{\rm f}=15$ days, and finally observes totally $n_{\rm f}=12$ independent 
fields in 6 months.
The time-ordered visibility samples are then compressed as described in $\S3.2$
to the pixelized visibility data set with $12 \times 12$ mosaic format and with
$10\arcmin$ pixel size. 
The pixel smoothing effect is significant only in the scan direction. 
In this situation $\tW_{\rm p}(\bv)$ in equation (38) is top-hat smoothing 
filter 
\be
   \tW_{\rm p}(v_x) = {{\sin(\pi v_x \Delta\theta_{\rm p})} \over 
               {\pi v_x \Delta\theta_{\rm p}}},
\ee
where $\Delta\theta_{\rm p} = 10\min$ is the real-space pixel size used 
in the data compression, and $x$ denotes the scan direction.
The visibility covariance matrix for the top-hat filtered drift-scan samples 
is obtained by modifying equations (6) and (7),
\be
   M_{mn}^{ij} = {\partial B_{\nu_i} \over \partial T} {\partial B_{\nu_j} 
      \over \partial T}
      \int d^2 w \tA_{\by_m}(\bu_i - \bw,\nu_i) \tA_{\by_n}^{*}
      (\bu_j-\bw, \nu_j) \tW_{\rm p}(u_{ix}-w_x)\tW_{\rm p}(u_{jx}-w_x)
      \mathcal{S}_{XY}(\bw),
\ee
and likewise for $N_{mn}^{ij}$.
The band power estimates from the visibilities obtained by the mock 
drift-scanning are shown in Figures 6$d$, 6$e$ and 6$f$, with a band width 
$\Delta u = 62.5$ except for the first and the last bands. 
The band power correlations between neighboring bands are from $-12$\% to 
$-10$\%.

\section{Fast Unbiased Estimator}

The AMiBA is expected to observe the sky over 100 deg$^2$ with the 
drift-scanning strategy. Even if the pixelization is applied to compress the
time-ordered visibility samples, it is still a formidable task for a data 
pipeline using the maximum likelihood quadratic estimator to compute 
the power spectra. In this section we propose a fast power spectrum estimator 
that can be implemented without recourse to large computing resource. 
The basic idea comes from the correlation function analysis, first developed 
by Szapudi et al. (2001), and from the MASTER method of Hivon et al. (2002). 
The visibilities are, by definition, the beam-convolved CMB Fourier modes 
(e.g., $V^{T}(\bu) \propto \int d^2 v \tA(\bv)\Delta\tT(\bu-\bv)$). 
Therefore, the correlation between a pair of visibilities are directly 
related with the power spectrum (Eqs. 6 and 7). 
We introduce an approximate expression,
\be
   V_i V_j - N_{ij} \approx S_{ij} = \sum_{b} B_{ij}(b) \pow_b,
\ee
where $B_{ij}(b) \equiv {{\partial S_{ij}} / {\partial \pow_b}}$
($i,j=1,2,\cdots,N_p$; $b=1,2,\cdots,N_b$).
The ensemble average of the above equation restores equation (11), 
$\left< V_i V_j \right> = S_{ij} + N_{ij}$. 

First, we perform the whitening transformation of the noise covariance matrix 
$\bN$, the instrument filter function $B_{ij}(b)$, and the visibility data 
vector $\bV$ as shown in equation (18).
Since $\bN$ is assumed to be diagonal, the transformed quantities are 
$\delta_{ij}$, $B_{ij}/\sigma_i\sigma_j$ ($\equiv B_{ij}^w$), 
and $V_i /\sigma_i$ ($\equiv V_i^w$), respectively. Equation (41) then becomes
\be
   V_i^w V_j^w - \delta_{ij} = \sum_b B_{ij}^w (b) \pow_b.
\ee
For a given band, complex pair weighting might be required for each 
$ij$-pair because of the uneven $uv$-coverage. However, we adopt an uniform
weighting scheme with equal weights.

After a careful rearrangement of all components of
$V_i^w V_j^w - \delta_{ij}$, $B_{ij}^w (b)$, and $\pow_b$ into simpler forms,
$\{v_k\}$, $\{\beta_{k b}\}$, and $\{c_{b}\}$, respectively, by merging 
an $ij$-pair to an index $k$, we can rewrite equation (42) as 
${\mbox{\boldmath $v$}}={\mbox{\boldmath $\beta$}} {\mbox{\boldmath $c$}}$. 
The fast unbiased estimator (FUE) of the CMB power spectrum is obtained as
\be
   {\mbox{\boldmath $c$}} 
   = \left[{\mbox{\boldmath $\beta$}}^T {\mbox{\boldmath $\beta$}}\right]^{-1} 
     {\mbox{\boldmath $\beta$}}^T {\mbox{\boldmath $v$}},
\ee
where 
\be
   \left({\mbox{\boldmath $\beta$}}^T {\mbox{\boldmath $\beta$}}\right)_{bb'} 
   = \sum_{i,j} {B_{ij}^w(b) B_{ij}^w(b')},
\ee 
and
\be
   \left( {\mbox{\boldmath $\beta$}}^T {\mbox{\boldmath $v$}}\right)_b 
         = \sum_{i,j}
           B_{ij}^w (b) \left( V_i^w V_j^w - \delta_{ij} \right).
\ee
The computation of equations (44) and (45), which are $\mathcal{O}(N_p^2)$ 
operations, is straightforward without need of any large memory and disk space.
The FUE method is fast especially when the noise covariance matrix $\bN$ 
is diagonal. Even for the non-diagonal noise covariance matrix, however,
the FUE is still fast since the prewhitening transformation is required 
just once.
If the visibility samples are sufficiently separated so that they are
considered as almost independent, the signal covariance matrix $\bS$ (also
$B_{ij}(b)$) can be approximated to be diagonal. This approach, which is very 
similar to a multi-band generalization of the Boughn-Cottingham statistic 
(Boughn et al. 1992), can accelerate the FUE speed from 
$\mathcal{O}(N_p^2)$ to $\mathcal{O}(N_p)$ operations.

To constrain cosmological models using the FUE estimated power spectrum, 
we need to know the uncertainty for each $\pow_b$ and the covariance 
between the band powers or the band power window functions.
The uncertainty of each band power estimated by FUE can be obtained by 
a Monte Carlo simulation (see, e.g., Hivon et al. 2002).
By fitting or interpolating the band power estimates $\{ \pow_b \}$, 
we can obtain a smooth CMB power spectrum $\pow_{\ell}^{\rm fit}$, 
from which many mock observations are made by including CMB signals,
instrumental noises, and other characteristics of the experiment.
The mock data sets are analyzed in the same way as the real data, 
giving rise to a set of power spectrum estimates
$\left\{ \{\pow_b^{\rm m_1}\}, \{\pow_b^{\rm m_2}\},\ldots \right\}$. 
The covariance matrix of the band powers 
\be
   C_{bb'} \equiv \left< 
     \left( \pow_b^{\rm fit}-\left<\pow_b^{\rm m}\right>_{\rm mock} \right) 
     \left( \pow_{b'}^{\rm fit}-\left<\pow_{b'}^{\rm m}\right>_{\rm mock} 
     \right) 
     \right>_{\rm mock}, 
\ee
is calculated.
The uncertainty for each band power $\pow_b$ is then given by the square
root of the diagonal components of $\bC$, i.e., 
$\Delta\pow_b = C_{bb}^{1/2}$. 

We apply the FUE method to the AMiBA 19-pointing mosaic data to measure 
the band powers $\{ \pow_b \}$ with error bars estimated from 30 simulation
data sets (open stars in Fig. 5). The band power estimates by FUE are quite 
consistent with those obtained by the quadratic estimator.
However, most of the FUE error bars are bigger than those of the quadratic 
estimator. 
Note that the FUE is a sub-optimal quadratic estimator with $\bN^{-1}$ 
compared to $(\bS+\bN)^{-1}$ of the optimal quadratic estimator.

\section{Discussion}

We have simulated interferometric observation of CMB temperature and 
polarization fluctuations.
For each observational strategy the data pipelines from the time-ordered raw 
visibility samples to the CMB angular power spectra ($\pow_b^{TT}$, 
$\pow_b^{EE}$, and $\pow_b^{TE}$) have been developed.
The pipelines are composed of making mock observation, data compression, and
power spectrum estimation. Data compression is achieved by pixelization 
of time-ordered visibilities in real- and $uv$-spaces by means of a common 
map-making process. This method can be applied to any kind of interferometric 
observation (see $\S3.2$).
In estimating the band powers from the mock visibility samples, the optimal
subspace filtering or signal-to-noise eigenmode analysis along with the 
quadratic estimator was used. By discarding the modes with low
signal-to-noise ratios, we were able to reduce the data set to a manageable 
size. One drawback of the optimal subspace filtering is that while it conserves
the information with signal-to-noise ratio higher than the limit of eigenvalue 
threshold, some useful information may be lost in certain 
cases. For instance, in measuring the CMB polarization power spectrum,
if the band width is too small to keep sufficient amount of the signal 
compared with the noise level, the weak signal can disappear during 
the optimal subspace filtering. Therefore, we need to choose a wider band 
width to obtain a higher signal-to-noise ratio, especially at higher 
$\ell$ region.

The measured band powers are found to be quite consistent with the band power 
expectation values $\left< \pow_b \right>$ for the AMiBA 19-pointing mosaic
(Fig. 5). This implies that our data pipelines are working reliably. 
Using the fact that the visibility contains direct information of CMB 
power spectrum, we have developed a fast unbiased estimator of the CMB 
power spectra (FUE, $\S5$) that requires only $\mathcal{O}(N_p^2)$ operations.
This method is very similar to the power spectrum estimation method using
Gabor transform (Hansen, G\'orski, \& Hivon 2002).
The FUE also gives band power estimates that are consistent with those 
from the quadratic estimator (see Fig. 5). The FUE method does not 
require large computer resources. Given the precomputed quantities 
$B_{ij}(b)=\partial S_{ij}/\partial \pow_b$, the computational speed 
is extremely fast. Even if the noise covariance matrix is highly non-diagonal,
which is the usual case in real data analyses (e.g., handling constraint 
matrices to subtract the point source effect), the FUE method is still fast 
because the prewhitening transformation of $B_{ij}(b)$ and $V_i$ is needed 
only once.

Our main goal was to propose data analysis techniques for each observational
strategy of a CMB interferometer, especially when the $uv$-beam size is larger
than the scale of structures in the CMB power spectrum.
Using the mock CBI single pointing observations, we have investigated
the effect of rotation of the array platform on the band power correlations 
and the uncertainties of the band powers.
Based on the results, summarized in Figure 2 and Table 2, we conclude that 
the band power anti-correlations can be reduced by rotating the platform 
and thus densely sampling the visibility plane. However, the uncertainties 
of the band power estimates slightly increase (when the total integration time 
is fixed). This is because the CMB signal is shared by the neighboring 
visibilities due to the finite beam size.
In this way, we can increase the resolution of the power spectrum 
in the $\ell$-space down to a resolution limit $\Delta\ell \approx \pi
/\theta$ given by the sampling theorem. 

Using the recent CBI result of single pointing observation, Mason et al. 
(2002) have shown a power spectrum with band width of 
$\Delta\ell = 4\sqrt{2} \ln 2 / \theta_{\rm fwhm} \approx 300$. 
This limit for $\Delta\ell$ is the FWHM of the visibility window function, 
which is proportional to the square of the Fourier transform of the primary 
beam with $\theta_{\rm fwhm}$ (Pearson et al. 2002).  
On the other hand, our choice for the band width is $\Delta\ell = 226$ 
($\Delta u = \Delta u_{\rm fwhm}/2 =36$). This is the limit given by
the sampling theorem ($\Delta\ell$ $\approx \pi/\theta$). 
It is $\Delta\ell = 4 \ln 2 / \theta_{\rm fwhm}$ for a Gaussian primary beam, 
which is a factor of $\sqrt{2}$ narrower than that adopted by CBI team.
We show in Table 2 that a mock CBI observation with 30 different orientations
results in about $20$\% anti-correlations between neighboring bands 
at $\ell \lesssim 1,000$, and higher values at higher $\ell \gtrsim 1,000$,
while they are 10 -- 15\% in Mason et al. (2002) due to the wider band
width. The band power correlation at high $\ell$ regions can be reduced 
by more densely sampling visibilities with sufficient integration time. 
As shown in the example of $uv$-mosaicking using a two-element interferometer
(Fig. 3), the band widths of power spectrum can be reduced while keeping
the band correlations at a tolerable level by increasing the number of rotation
steps with increasing dish separation, and by assigning longer integration time 
to the visibilities at low CMB signal regions. 
For intermediate five bands, the band width and the average band power 
correlation are $\Delta\ell = 132$ ($\Delta u =21$) and $-24$\%, respectively.
The width is smaller than our resolution limit ($\Delta\ell= 161$, 
$\Delta u = \Delta u_{\rm fwhm}/2$ where $\Delta u_{\rm fwhm} = 51.2$), also
a factor of 1.7 narrower than the limit obtained by Pearson et al.'s formula
($\Delta\ell \simeq 230$).

The recent DASI power spectrum is measured from the single pointing
observation without platform rotation (Halverson et al. 2002). 
The DASI band powers have the resolution of $\Delta\ell \approx 80$
(with 18 -- 28\% anti-correlations), which is broader than the 
resolution limit $\Delta\ell = 4\sqrt{2} \ln 2 / \theta_{\rm fwhm}= 66$ 
where $\theta_{\rm fwhm} = 3\fdg4$. 
We expect that the DASI single pointing observation with dense rotation 
of platform will allow higher resolution of about $\Delta\ell 
= 50$ at the similar level of anti-correlations.
Since the mosaicking is the most efficient method for increasing the resolution
of the power spectrum, the combination of mosaicking and dense rotation 
of the platform followed by the $uv$-pixelization is thought to be the most 
ideal observational strategy for DASI- and CBI-type CMB interferometers.

For each observational strategy, optimal parameter choices for the AMiBA
experiment are discussed in $\S4$, and summarized in Table 3.
The 7-element AMiBA is expected to detect CMB polarization power spectrum
near $\ell \approx 1,300$ at $4\sigma$ level within 20 days by observing 
43 fields. In AMiBA mosaicking with $t_{\rm tot}=6$ months,
the optimal parameter sets are ($\theta_{\rm opt}=7\deg$, $n_{\rm f}=1$) or 
($\theta_{\rm opt}=3\deg$, $n_{\rm f}=5$) for a minimum uncertainty of
the $E$-polarization power spectrum.
In fact, the optimal parameters strongly depend on the characteristics 
of the interferometer (e.g., $\eta_{s}$, $\eta_{a}$, and $T_{sys}$) and 
on the $E$-polarization power spectrum $\pow_b^{EE}$. 
Since we are considering the shortest baselines in deriving the parameters, 
the optimization is only for the sensitivity range of the shortest baselines 
($\ell < 2,000$). At higher $\ell$-range ($\ell > 2,000$) where the 7-element 
AMiBA has only a few baselines, the CMB polarized signal is expected to be
very low. Therefore, we have chosen a wider band width for the last band for 
the $12 \times 12$ mosaicking and drift-scanning observations (see Fig. 6). 
To obtain a meaningful polarization power spectrum at high $\ell$
region with narrow band widths, we need to increase the integration time or
the number of baselines.
This can be seen in the simulation of the 19-pointing mosaicking by AMiBA 
where the integration time per pointing is almost one day (see Fig. 5).
Although the band widths are quite wide in the temperature power spectra
in Figure 6, we can measure temperature band powers independently 
with narrower band width because the signal-to-noise ratios of the $T$ 
visibilities are very high, compared with those of polarization 
($\Delta\ell = 196$, see open circles in Fig. 6$d$).

Among the three observational strategies that we have studied,
the single pointing is useful for a detection of the CMB polarized signal
while the mosaicking or the drift-scanning of a large area of the sky is
essential for measuring the polarization power spectrum with high 
$\Delta\ell$-resolution. The drift-scanning strategy is efficient for removing 
the ground contamination. It can also save half of the integration time 
when compared to the method of differencing two fields in the removal of 
the ground spillover adopted by the CBI experiment (Padin et al. 2001).
In the drift-scanning, the survey region can have a shape for which the 
flat-sky approximation is inapplicable. 
Since the survey area drift-scanned by the AMiBA interferometer will be over 
100 deg$^2$, it is necessary to take into account the curvature of the sky.
Our future work will deal with important issues such as the removal 
of Galactic foreground emission, the identification of radio point sources, 
and the subtraction of unresolved point sources. 
It is also important to study the topology of the CMB temperature and 
polarization fields to test the primordial fluctuations for Gaussianity  
(see, e.g., Park \& Park 2002).

\acknowledgments
We acknowledge valuable discussions with J.-H. Proty Wu, Ue-Li Pen, 
Tzihong Chiueh, Mike Kesteven, Cheng-Juin Ma, and helpful comments from the
anonymous referee. 
CGP and CP were supported by the BK21 program of the Korean Government
and the Basic Research Program of the Korea Science \& Engineering Foundation
(grant no. 1999-2-113-001-5). 
CGP acknowledges helpful comments with Hyun Seok Yang, and the support 
by the CosPA Center at the National Taiwan University during his visit. 
KWN was supported in part by the National Science Council, Taiwan ROC 
under the Grant NSC90-2112-M-001-028.

\clearpage

\begin{deluxetable}{cc}
\tablewidth{0pt}
\tablecaption{
Ensemble Averages of the Conjugate Products of $T$, $Q$, and $U$ Pairs 
in the $uv$-Plane.
}
\tablehead{
\colhead{$XY$} &
\colhead{$\left<\tX(\bv)\tY^{*}(\bw)\right> = \mathcal{S}_{XY}(\bw) 
          \delta(\bv-\bw)$} 
}
\startdata
$TT$  & $\mathcal{S}_{TT}(\bw)= S_{TT}(w)$ \\
$TQ$  & $\mathcal{S}_{TQ}(\bw)= S_{TE}(w) \cos 2\phi_{\bw}$ \\
$TU$  & $\mathcal{S}_{TU}(\bw)= S_{TE}(w) \sin 2\phi_{\bw}$ \\
$QQ$  & $\mathcal{S}_{QQ}(\bw)= S_{EE}(w) \cos^2 2\phi_{\bw} + S_{BB}(w) \sin^2 2\phi_{\bw}$ \\
$QU$  & $\mathcal{S}_{QU}(\bw)= S_{EE}(w) \cos 2\phi_{\bw} \sin 2\phi_{\bw}
         - S_{BB}(w) \sin 2\phi_{\bw} \cos 2\phi_{\bw}$ \\
$UU$  & $\mathcal{S}_{UU}(\bw)= S_{EE}(w) \sin^2 2\phi_{\bw} + S_{BB}(w) \cos^2 2\phi_{\bw}$ \\
\enddata
\end{deluxetable}

\clearpage

\begin{deluxetable}{ccccccccccc}
\tablecolumns{11}
\tablewidth{0pt}
\tablecaption{
Band Power Correlations $M_{bb'}/\sqrt{M_{bb} M_{b'b'}}$ and Relative 
Uncertainties $\Delta\pow_b / \pow_b$ in the Mock CBI Observations\tablenotemark{a}.}
\tablehead{
\colhead{} & \colhead{} & \multicolumn{9}{c}{$M_{bb'} /\sqrt{M_{bb} M_{b'b'}}\tablenotemark{b}$} \\ 
\cline{3-11} 
\colhead{$n_{\rm f}$}  &
\colhead{$t_{\rm f}$}  & 
\colhead{$(b,b')$}  &
\colhead{$(1,2)$}  &
\colhead{$(2,3)$}  &
\colhead{$(3,4)$}  &
\colhead{$(4,5)$}  &
\colhead{$(5,6)$}  &
\colhead{$(6,7)$}  &
\colhead{}         & 
\colhead{Average}  
}
\startdata
$30$ & $72^{\rm h}$ && $-0.35$ & $-0.34$ & $-0.34$ & $-0.43$ & $-0.48$ & $-0.57$ & \nodata & $-0.41$  \\
$15$ & $24^{\rm h}\times 6^{\rm ori.}$ && $-0.25$ & $-0.29$ & $-0.27$ & $-0.40$ & $-0.44$ & $-0.43$ & \nodata & $-0.35$  \\ 
$9$ & $24^{\rm h} \times 10^{\rm ori.}$ && $-0.21$ & $-0.23$ & $-0.23$ & $-0.33$ & $-0.40$ & $-0.44$ & \nodata & $-0.30$  \\
$6$ & $12^{\rm h} \times 30^{\rm ori.}$ && $-0.21$ & $-0.21$ &  $-0.22$  & $-0.31$  &  $-0.33$ &  $-0.36$  & \nodata & $-0.27$   \\
\hline\hline
\colhead{} & \colhead{} & \multicolumn{9}{c}{$\Delta\pow_b / \pow_b$} \\
\cline{3-11} 
$n_{\rm f}$ & $t_{\rm f}$ & $b$ & $1$ & $2$ & $3$ & $4$ & $5$ & $6$ & $7$ & Average \\
\cline{1-11}
$30$ & $72^{\rm h}$ && 0.15 & 0.23 & 0.11 & 0.17 & 0.14 & 0.18 & 0.21 & 0.17 \\
$15$ & $24^{\rm h}\times 6^{\rm ori.}$ && $0.15$ & $0.27$ & $0.14$ & $0.17$ & $0.20$ & $0.18$ & $0.19$ & $0.19$\\
$9$ & $24^{\rm h} \times 10^{\rm ori.}$ && 0.14 & 0.28 & 0.15 & 0.20 & 0.21 & 0.22 & 0.25 & 0.21 \\
$6$ & $12^{\rm h} \times 30^{\rm ori.}$ && 0.13 & 0.33 & 0.20 & 0.23 & 0.24 & 0.22 & 0.25 & 0.23 \\
\enddata
\tablenotetext{a}{The total observation time is set to $t_{\rm tot}=90$ days.}
\tablenotetext{b}{$M_{bb'}$ is the inverse Fisher matrix element 
$(\bF^{-1})_{bb'}$.}
\tablecomments{Only the first 7 band power estimates ($\Delta u=36$ or
$\Delta\ell = 226$) are considered.}
\end{deluxetable}

\clearpage

\begin{deluxetable}{ccccc}
\tablecolumns{5}
\tablewidth{0pt}
\tablecaption{
Optimal Parameters for AMiBA Single Pointing and Mosaicking 
Strategies\tablenotemark{a}.
}
\tablehead{
\multicolumn{5}{c}{Single Pointing} 
}
\startdata
\colhead{$n_d$\tablenotemark{b}} & \colhead{} & \colhead{$n_{\rm f}$} & 
\colhead{$t_{\rm f}$} & \colhead{$\Delta\pow_b$\tablenotemark{c}} \\
\cline{1-1} \cline{3-5}
2 &  & 32 & 135 hrs & 10 $\uK^2$ \\
7 &  & 380 & 11.4 hrs & 2 $\uK^2$ \\
\hline\hline 
\multicolumn{5}{c}{Square Mosaicking ($n_d=7$)}\\
\hline
\colhead{$\theta_{\rm opt}$} & \colhead{} & \colhead{$n_{\rm f}$} &
\colhead{$t_{\rm f}$} & \colhead{$N_{\rm mo}$ ($\delta\theta_{\rm mo}=15\arcmin$)} \\
\cline{1-1} \cline{3-5}
$7\deg$  &  & 1 & 180 days & $28\times 28$ \\
$3\deg$  &  & 5 & 36 days & $12\times 12$ \\
\enddata
\tablenotetext{a}{The total observation time is set to $t_{\rm tot}=6$ months.}
\tablenotetext{b}{Number of dishes.}
\tablenotetext{c}{$E$-mode band power uncertainty. The band width 
of the AMiBA shortest baselines is $\Delta u=154$, centered at $u_b=206$.}
\end{deluxetable}

\clearpage

\begin{figure}
\resizebox{\textwidth}{!}{\includegraphics{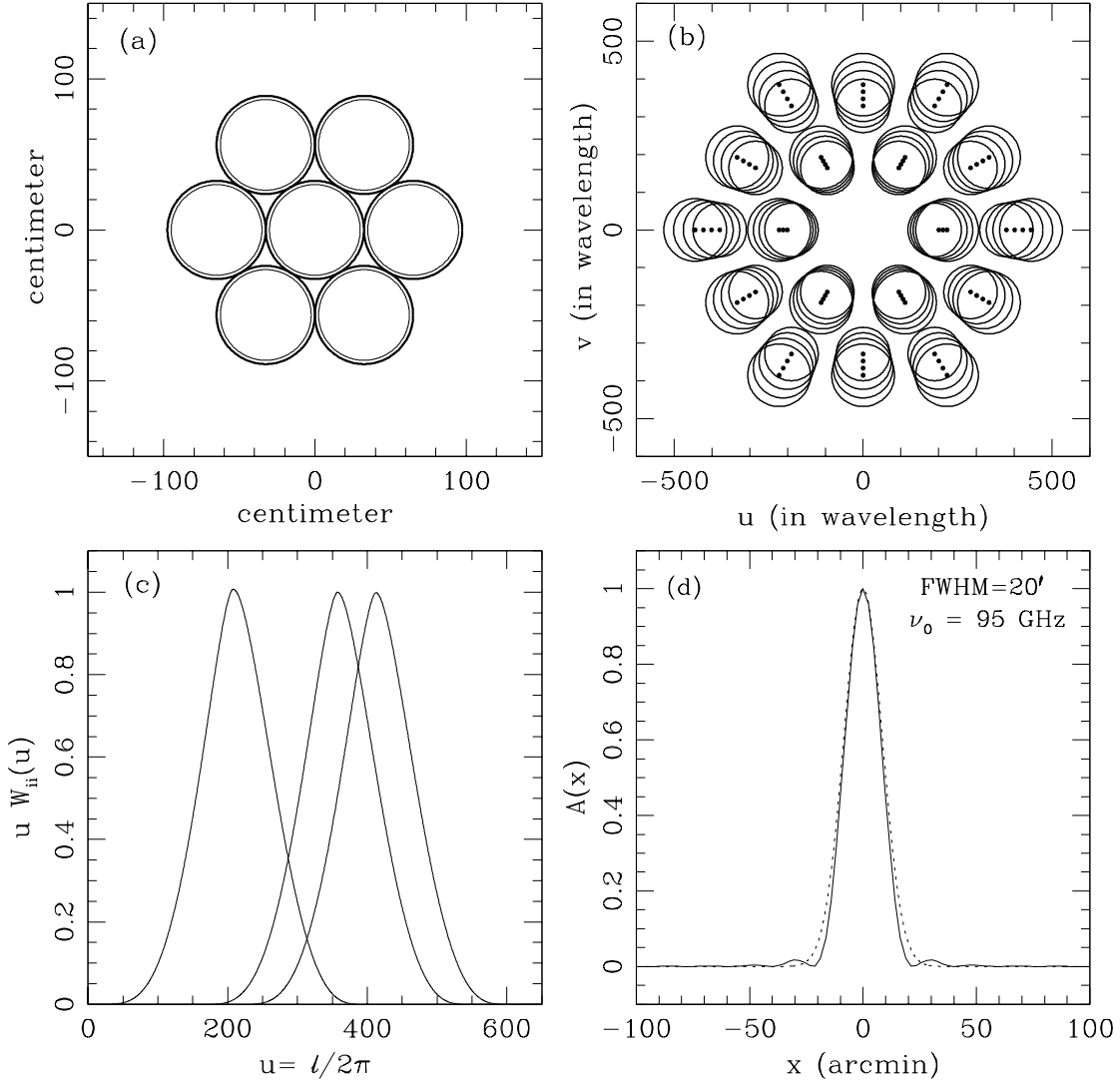}}
\caption{($a$) Close-packed hexagonal configuration of seven receivers
($D=60$ cm) of the AMiBA on a single platform. Each feed horn can detect
$T$, $Q$, $U$, and $V$ Stokes parameters simultaneously using complex dual
polarizer with $T_{sys}=70$ K. 
Thin circles are boundaries of the physical dishes, and the shields 
between the 5 cm gaps of the adjacent dishes are drawn as thick circles. 
Due to the hexagonal configuration, the instrument gives 21 baselines with 
3 different lengths in different directions. 
($b$) The $uv$-coverage for a snapshot of the sky with the configuration shown 
in ($a$). Four frequency channels are assumed ($\nu_0 = 95$ GHz, $\Delta\nu=20$
GHz whole bandwidth). The points are the $uv$-coordinates of the baseline 
vectors, and the circles have diameters equal to the FWHM of the $uv$-beam 
pattern.
($c$) The sensitivity of AMiBA at the center frequency $\nu_0 = 95$ GHz
shown by the window functions for the three baselines with different lengths
(Eq. 14 of W99).
($d$) The real-space primary beam pattern $A(\bx)$ at the center frequency 
(solid curve), and a Gaussian function with ${\rm FWHM}\simeq 20\min$ 
(dotted curve) approximating the beam.} 
\end{figure}
\clearpage

\begin{figure}
\epsscale{1.5}
\resizebox{\textwidth}{!}{\includegraphics{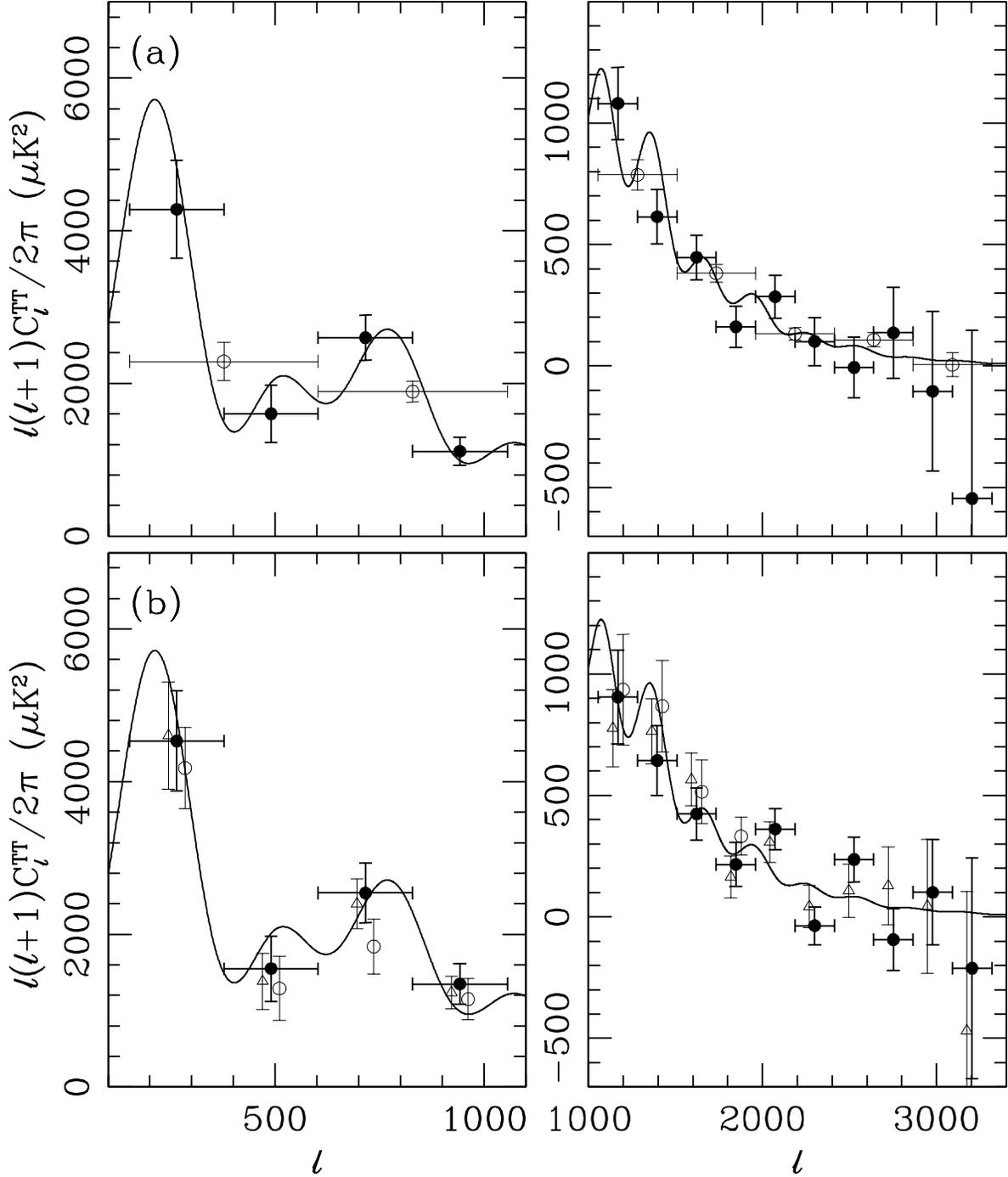}}
\caption{($a$) The power spectra of the CMB temperature anisotropy measured 
from a mock CBI observation ($n_{\rm f}=30$, $t_{\rm f}=72^{\rm h}$) with band 
power widths $\Delta u=36$ (filled circles) and $72$ (open circles).
($b$) Power spectra measured from mock CBI observations with 6 ($n_{\rm f}=15$,
$t_{\rm f}=24^{\rm h} \times 6$; open triangles), 10 ($n_{\rm f}=9$,
$t_{\rm f}=24^{\rm h} \times 10$; filled circles), and 30 
($n_{\rm f}=6$, $t_{\rm f}=12^{\rm h} \times 30$; open circles) 
orientations. 
Each figure is divided into two panels at low and high $\ell$ regions. 
To avoid confusion, the open triangles and circles are slightly 
shifted to the left and right with respect to the filled circles, and 
open circles at $\ell > 2000$ are omitted.}
\end{figure}
\clearpage

\begin{figure}
\epsscale{1.5}
\resizebox{\textwidth}{!}{\includegraphics{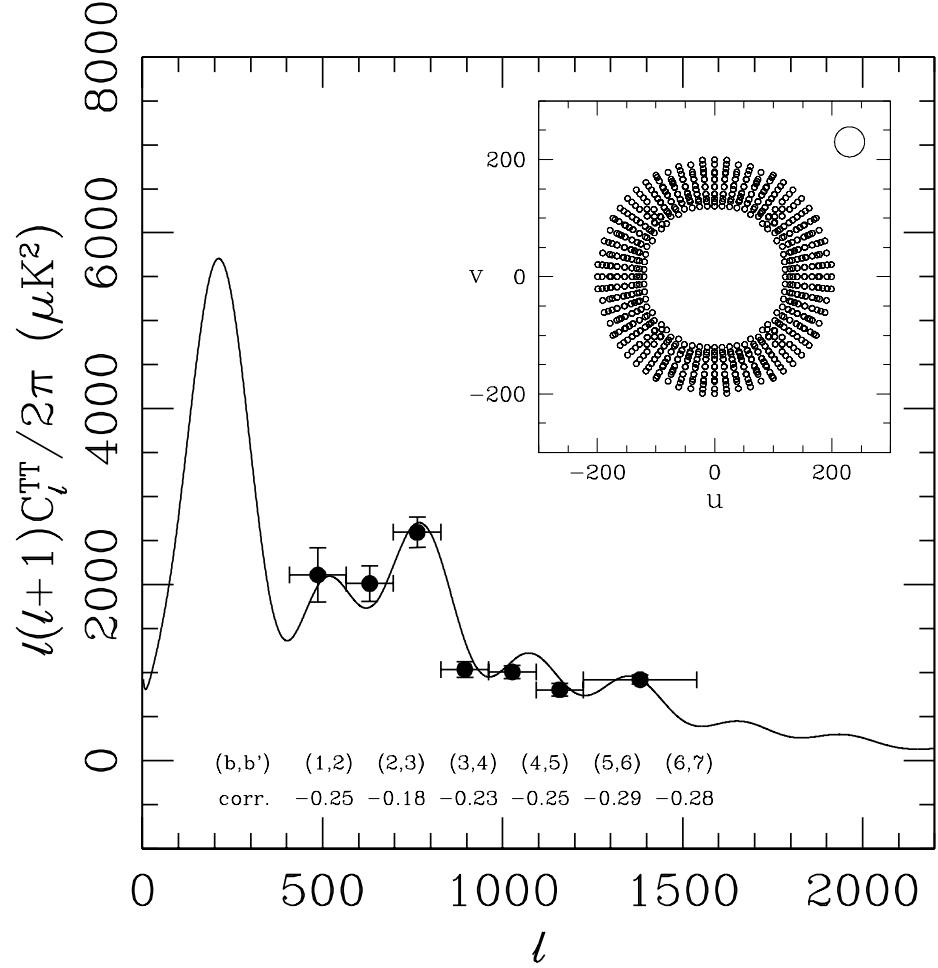}}
\caption{The CMB temperature anisotropy power spectrum estimated from 
30 mock observations by a two-element interferometer ($D=20$ cm, two channels 
with $\Delta\nu=20$ GHz whole bandwidth centered at 95 GHz). 
The dish separation changes from 40 cm to 60 cm, and the platform
is rotated to obtain uniform visibility samples. The distribution
of visibilities in the $uv$-plane is shown in the upper right box. 
In the upper right corner of the box, the primary beam is shown as a circle 
with diameter $\Delta u_{\rm fwhm}=51.2$.
Except for the first ($\Delta u = 25$) and the last ($\Delta u =50$) bands, 
the band width of the intermediate five bands is chosen to be 
$\Delta u = 21$ ($\Delta\ell = 132$). 
The band power correlations between neighboring bands are also given at the 
bottom of the figure.
}
\end{figure}
\clearpage

\begin{figure}
\resizebox{\textwidth}{!}{\includegraphics{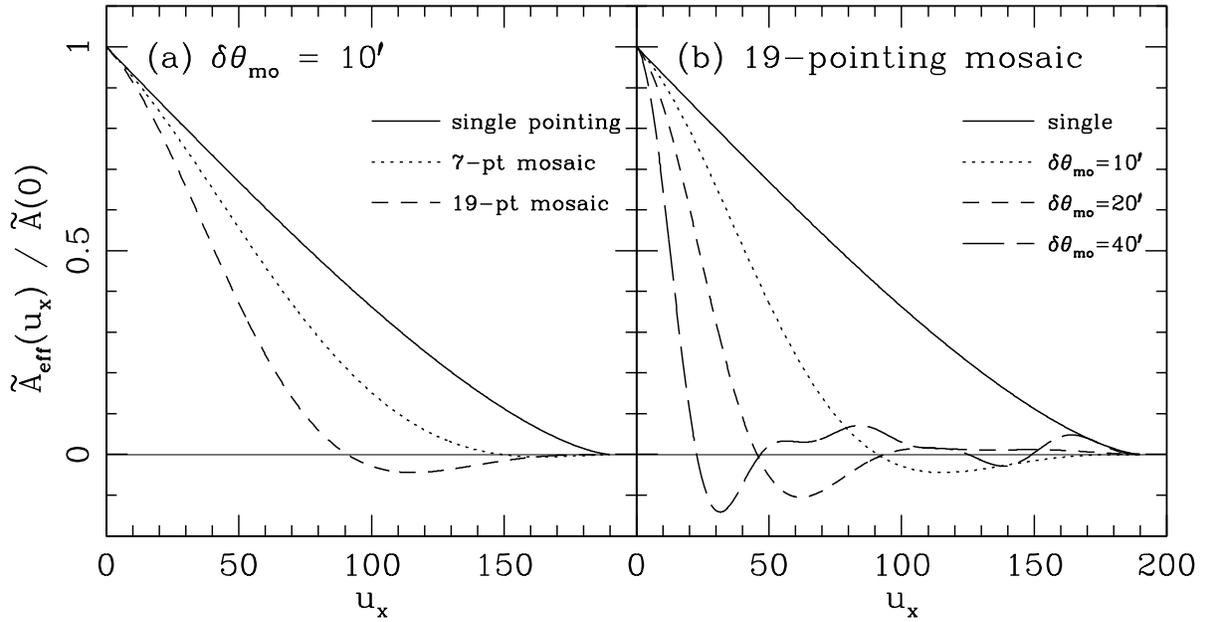}}
\caption{Effective beam patterns expected ($a$) in a single pointing 
observation (solid curve) and 7- and 19-point hexagonal mosaicking 
(dotted and dashed curves) with pointing separation 
$\delta\theta_{\rm mo}=10\arcmin$, and ($b$) in 19-point hexagonal mosaics 
with pointing separations of $\delta\theta_{\rm mo}=10\arcmin$, 
$20\arcmin$, and $40\arcmin$. The pattern of the hexagonal mosaic is similar 
to that of the AMiBA dishes shown in Figure 1$a$.
Since the hexagonal configuration makes the imaginary components of the 
individual beam patterns along the $u_x$ and $u_y$ axes cancel 
with one another, the effective beam patterns are real functions. 
}
\end{figure}
\clearpage

\begin{figure}
\resizebox{\textwidth}{!}{\includegraphics{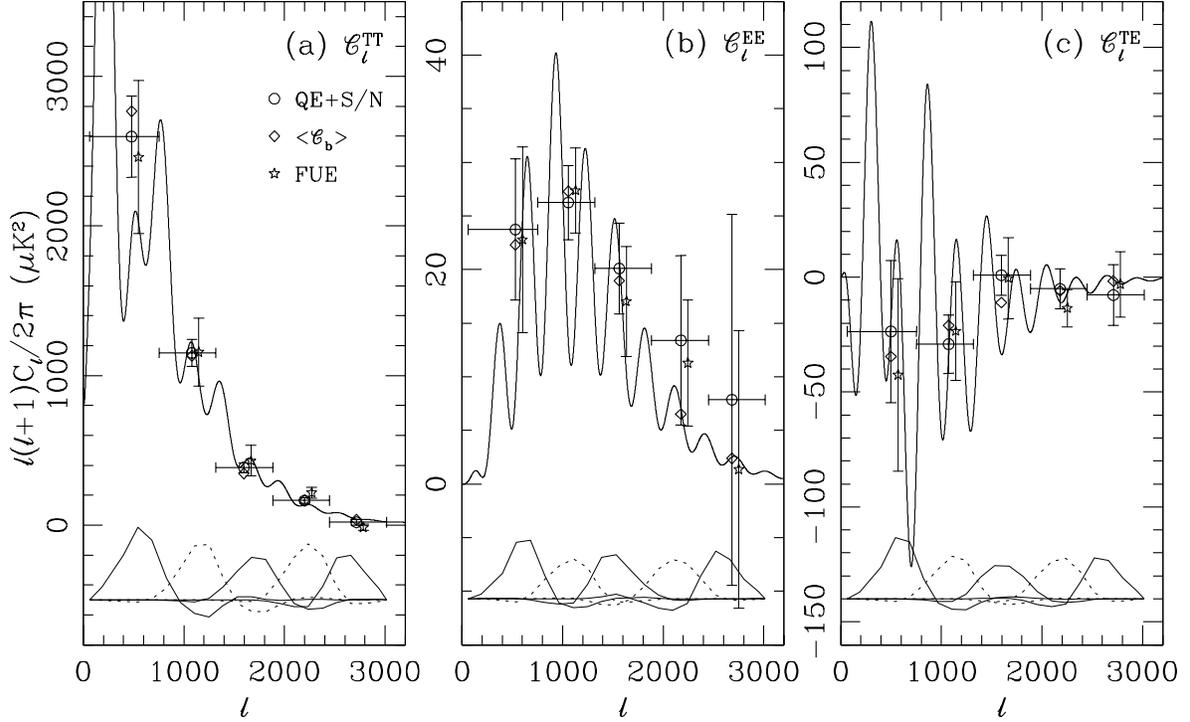}}
\caption{Band power estimates for ($a$) the temperature,
($b$) $E$-polarization, and ($c$) $TE$ cross-correlation power spectra 
measured from the AMiBA 19-pointing hexagonal mosaic mock observation 
($n_{\rm f}=10$, $t_{\rm f}=22\fh7 \times 19$, and 
$\delta\theta_{\rm mo}=15\arcmin$). Open circles with error bars
are the results of the quadratic estimator (QE) and the signal-to-noise 
eigenmode analysis (S/N).
For each band, the band power expectation value $\left< \pow_b \right>$
(denoted by an open diamond) and the corresponding band power window function
$f_\ell^b$ $\equiv W_\ell^b / \ell$ (denoted by a solid or a dotted curve 
in the bottom region) are also shown.
The $\ell$-location of each band power ($\ell_{\rm eff}$) is found from 
$\ell_{\rm eff} = \sum_{\ell} \ell f_\ell^b / \sum_{\ell} f_\ell^b$
(Bond et al. 1998).
The band power estimates obtained by the fast unbiased estimator (FUE), 
right-shifted with respect to $\ell_{\rm eff}$ and denoted by open stars, 
are consistent with those of the quadratic estimator. 
The FUE error bars are estimated from 30 simulations (see $\S5$).
}
\end{figure}
\clearpage

\begin{figure}
\resizebox{\textwidth}{!}{\includegraphics{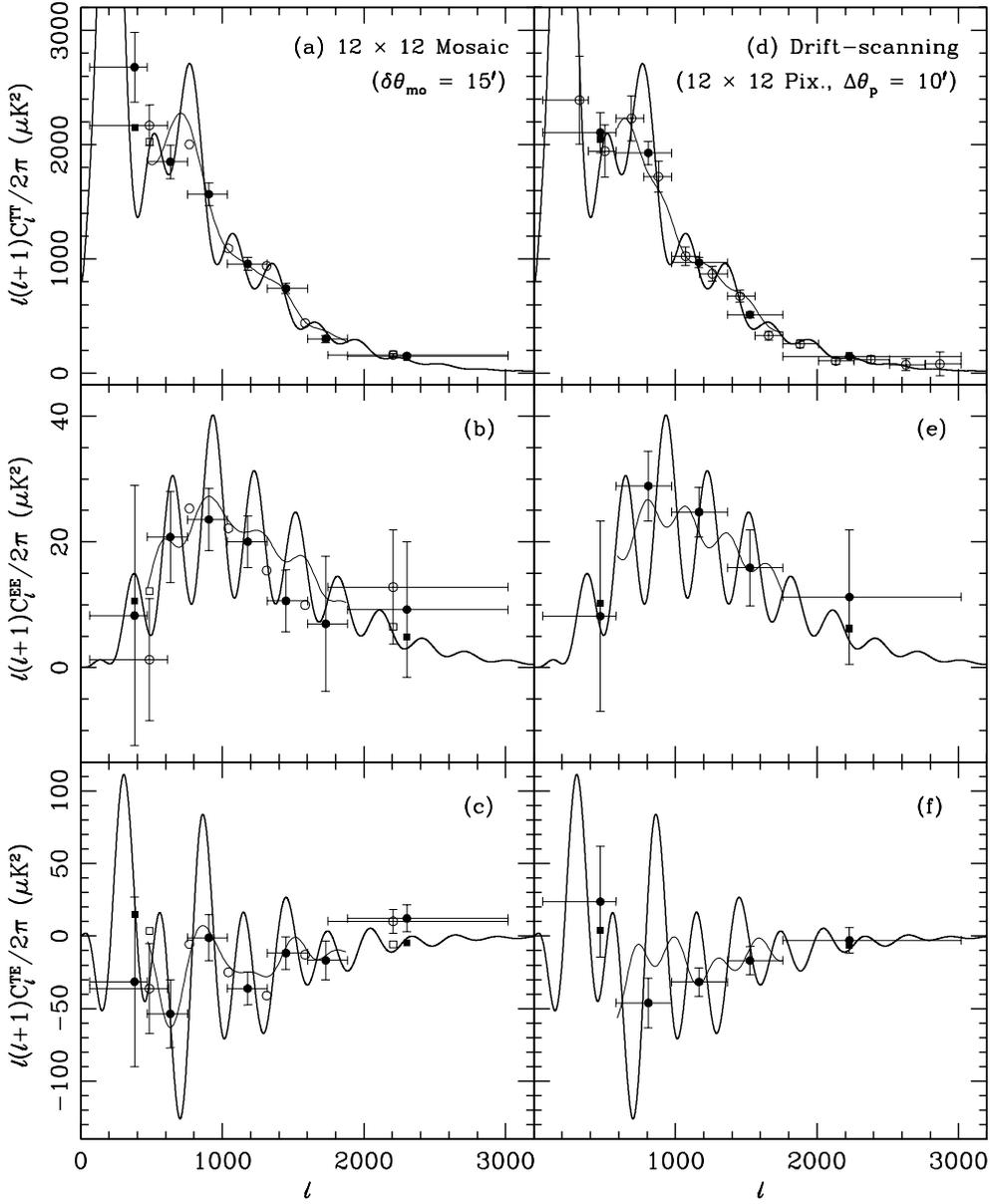}}
\caption{($a$) Temperature, ($b$) $E$-polarization, 
and ($c$) $TE$ cross-correlation power spectra measured from an AMiBA 
$12\times 12$ mosaicking observation with $\theta_{\rm opt}=3\deg$, 
$n_{\rm f}=5$, and $\delta\theta_{\rm mo} = 15\arcmin$ for 6 months.
Band powers are measured at two sets of band centers (filled and open circles)
shifted by a half of the band width with respect to each other.
Band power estimates from the data obtained by the AMiBA drift-scanning 
strategy ($\theta_{\rm opt}=2\deg$, $n_{\rm f}=12$, and $t_{\rm tot}=6$ months)
followed by the $12\times 12$ pixelization with $\Delta\theta_{\rm p} = 
10\arcmin$ are shown in ($d$), ($e$) and ($f$).  
Open circles in the panel ($d$) are the temperature power spectrum measured 
with the band width ($\Delta u = 31.25$) twice smaller than that of the filled 
circles. The location of each band power is estimated using the window function
defined in equation (37). 
Thick curves are the $\Lambda$CDM model power spectra. The model power spectra
smoothed by a box of band power width ($\Delta u = 45$ for mosaicking, and 
$\Delta u = 62.5$ for drift-scanning) weighted by the window function  
are also shown (thin curves, and filled and open squares).
}
\end{figure}
\clearpage

\end{document}